\documentclass[prd,aps,amsfonts,eqsecnum,superscriptaddress,nofootinbib,longbibliography,notitlepage]{revtex4-1}

\usepackage{graphicx}
\usepackage{xcolor}
\usepackage{caption}
\usepackage{rotating}
\usepackage{amsmath,amssymb,graphics,amsthm,isomath}
\usepackage{subfigure}

\usepackage[colorlinks=true, urlcolor=violet, linkcolor=blue, citecolor=red, hyperindex=true, linktocpage=true]{hyperref}
\usepackage[capitalise,compress]{cleveref}

\numberwithin{thm}{section}

\makeatletter
\renewcommand{\p@subsection}{}
\renewcommand{\p@subsubsection}{}
\makeatother

\usepackage{xcolor}
\usepackage{mathtools}

\newcommand{\ii}{\mathrm{i}}

\usepackage{dsfont}

\begin{document}

\title{Infinite families of fracton fluids with momentum conservation}

\author{Andrew Osborne}
\email{andrew.osborne-1@colorado.edu}
\affiliation{Department of Physics and Center for Theory of Quantum Matter, University of Colorado, Boulder CO 80309, USA}

\author{Andrew Lucas}
\email{andrew.j.lucas@colorado.edu}
\affiliation{Department of Physics and Center for Theory of Quantum Matter, University of Colorado, Boulder CO 80309, USA}

\date{\today}

\begin{abstract}
  
  We construct infinite families of new universality classes of fracton hydrodynamics with momentum conservation, both with multipole conservation laws and/or subsystem symmetry.  We explore the effects of broken inversion and/or time-reversal symmetry at the ideal fluid level, along with momentum relaxation.  In the case of one-dimensional multipole-conserving models, we write down explicit microscopic Hamiltonian systems realizing these new universality classes.  All of these hydrodynamic universality classes exhibit instabilities and will flow to new non-equilibrium fixed points.  Such fixed points are predicted to exist in arbitrarily large spatial dimensions.
  
\end{abstract}
\maketitle

\tableofcontents

\section{Introduction}
Recent years have seen the discovery and classification of many new phases of quantum matter with elementary excitations, dubbed \emph{fractons}, which exhibit restricted mobility \cite{emergent_Nandkishore,cage_Hermele,pinch_Nandkishore,symmetric_Pretko,twisted_Delgado,fracton_Kim,quantum_Kim,xcube_Kim,foilated_Williamson,higher_Pretko,correlation_Sondhi,fractal_Sondhi,subsystem_You,symmetric_You,fractonic_You,absence_Weinstein,higher_Wang,field_Seiberg,topological_Aasen,fracton_Hermele,fractonic_Ye,topological_Nandkishore,recoverable_Nandkishore,higher_Pretko_2,thermalization_Bernevig,fractons_sous}.   More recently, it has been noticed that these same phases of quantum matter give rise to novel universality classes of hydrodynamic behavior \cite{fracton_hydro}, which arise due to the conservation of exotic space-dependent conserved charges, such as the total dipole moment of a system \cite{knap2020,morningstar,conservation_Nandkishore,hart2021hidden,sala2021dynamics}, or the charge along lines and/or planes in a higher-dimensional lattice \cite{IaconisVijayNandkishore,knap2021}. 

The first studies of such hydrodynamic behavior focused on the subdiffusion of a single conserved charge, which was immediately relevant for both numerical studies of random quantum circuits \cite{conservation_Nandkishore,knap2020} as well as tilted Fermi-Hubbard models in optical lattices \cite{Guardado_Sanchez_2020,zhang2020universal}.  However, more recently, it has also been noted that these fracton fluids exhibit interesting and unexpected behavior when momentum is also conserved \cite{lt4dbreakdown_Lucas,hydroscoop_Surowka}; in particular, hydrodynamics is generally unstable in physically realizable dimensions \cite{lt4dbreakdown_Lucas}.

In this paper, we will describe infinitely many new families of hydrodynamic models with ``fractonic" conservation laws, as well as momentum conservation.  Our goal is not to systematically analyze any one of them (as \cite{lt4dbreakdown_Lucas} did for the dipole-conserving fluid), but instead to draw some general lessons about the way in which momentum conservation can modify fracton hydrodynamics.  A detailed understanding of this issue will help lead to systematic field theories which couple fractonic models to gravity (if and when this is even possible \cite{Slagle:2018kqf}): after all, the effective field theories of hydrodynamics necessarily couple a momentum-conserving fluid to a spacetime  or vielbein (when classical background fields are accounted for) \cite{EFT_Liu}.  But as the field theoretic construction of a momentum-conserving universality class is quite non-trivial \cite{paolotoappear,Jain:2021ibh,Bidussi:2021nmp,Pena-Benitez:2021ipo}, we will instead seek to analyze our new universality classes using simpler methods grounded in classical Hamiltonian mechanics.   We expect that our constructions will help to make progress on these more challenging theoretical problems.

In Section \ref{ideal_hydro}, we use a continuum Hamiltonian formulation to describe fracton fluids with momentum conservation.  In Section \ref{towards_diss}, we will begin to discuss dissipative effects -- in particular, the consequences of breaking momentum conservation.  We will both recover more conventional fractonic subdiffusion, and also argue for novel universality classes that arise without time-reversal symmetry (which have not yet been discovered).  In Section \ref{microscopic}, we present one-dimensional discrete Hamiltonian models which can form the basis of large-scale numerical simulations for theories with quadrupole (and higher) conservation laws, and confirm that hydrodynamic quasinormal modes within ideal hydrodynamics match with our more generic predictions.


\section{Ideal hydrodynamics in the continuum}\label{ideal_hydro}
We begin by developing a simple continuum Hamiltonian formalism for hydrodynamics with fractonic conservation laws.   This approach is more conceptually clean and will naturally work in all spatial dimensions $d$.
\subsection{Poisson brackets}
Let $\rho(x)$ and $\pi_i(x)$ denote the charge density and $d$ components of momentum density.  We assume that the coordinates $x\in \mathbb{R}^d$ live on the plane.  
Following \cite{Son:2000ht}, we define a classical Poisson bracket which correctly incorporates the effects of translations on the classical fields:
\begin{subequations}\label{eq:brackets}\begin{align}
  \{\pi_i(x), \rho(y)\} &= \rho(x)\partial_i \delta(x - y), \label{eqn:pb} \\
  \{\pi_i(x), \pi_j(y)\} &= (\pi_j(x) \partial_i - \pi_i(x) \partial_j) \delta(x - y), \label{eqn:pb_cor} \\
  \lbrace \rho(x),\rho(y)\rbrace &= 0.
\end{align}\end{subequations}
Given some Hamiltonian $H$, we arrive at equations of motion by requiring that a variable $\phi$ evolves as 
\begin{equation}
  \partial_t \phi = \{\phi,H  \}.
\end{equation}
Spatial locality demands that the Hamiltonian $H$ be expressed as a local integral:\begin{equation}
    H = \int \mathrm{d}^dx \; \mathcal{H}(\rho, \pi_i),
\end{equation}
in terms of a Hamiltonian density $\mathcal{H}$.

It is very useful to define \begin{equation}
    v_i := \frac{\pi_i}{\rho}.
\end{equation}
Indeed, using (\ref{eq:brackets}), we notice that \begin{equation}
    \lbrace v_i(x), \rho(y)\rbrace = \partial_i \delta(x-y).
\end{equation}
Because the commutator of $v_i$ and $\rho$ leads to a field-independent object, we will find it most natural to write $\mathcal{H}$ in terms of $v_i$, rather than $\pi_i$.  The equations of motion for $\rho$ and $v_i$ then become particularly simple: 
\begin{subequations}\begin{align}\label{eqn:eomsfr}
    \partial_t \rho &= -\partial_i \frac{\delta H}{\delta v_i} \\
    \partial_t v_i &= -\partial_i \frac{\delta H}{\delta \rho} -\frac{1}{\rho} (\partial_i v_j - \partial_j v_i) \frac{\delta H }{\delta v_j}
\end{align}\end{subequations}
We will primarily focus on models where the latter term in the equation of motion for $v_i$ does not contribute. 

\subsection{Thermodynamics}
The requirement that our theory is consistent with thermodynamics leads to additional insight into the nature of these equations.  In particular, note that (ignoring temperature and \emph{thermal} hydrodynamic fluctuations, which will not play much of a role in this paper although certainly of importance in principle):
\begin{equation}\label{eqn:Hdiff}
  \mathrm{d}\mathcal{H} = - \mu \mathrm{d}\rho - V_i \mathrm{d}\pi^i = -\mu \mathrm{d}\rho - V_i (v_i \mathrm{d}\rho + \rho \mathrm{d}v_i) .
\end{equation}
We emphasize that $V_i$ is defined as the chemical potential of the $i$th component of momentum density, and that in general $V_i$ is \emph{very different} from $v_i$. 
Requiring that pressure $P$ is the Legendre transform of $H$, 
\begin{equation}\label{eqn:firstlaw}
  \mathrm{d}P = \mathrm{d}H + \mathrm{d}(\mu \rho) +\mathrm{d}(V_i\pi^i),
\end{equation}
we see that  
\begin{equation}\label{eqn:pdefn}
    \mathrm{d}P = \rho \mathrm{d}\mu + \pi_j \mathrm{d}V^j.
\end{equation}
Note that (\ref{eqn:firstlaw}) and  (\ref{eqn:pdefn}) simply represent our first law of thermodynamics, with the important caveat that we are not including energy density among the conserved modes to keep track of.\footnote{Alternatively, one may wish to consider states at constant entropy density, so that the $T\mathrm{d}s$ contribution to (\ref{eqn:Hdiff}) vanishes.}
It follows from (\ref{eqn:Hdiff})  that 
\begin{subequations}\label{eqn:Vdefn}
  \begin{align}
    \frac{\delta H}{\delta \rho} &= V_iv^i + \mu, \label{eqn:mudefn}\\
    \frac{\delta H}{\delta v_i} &= V^i \rho,  \label{eqn:Videfn}.
  \end{align}
\end{subequations}
These definitions are sufficient to show  
\begin{equation}
  -\partial_t \pi_i =  -\rho \partial_t v_i - v_i \partial_t \rho = \rho \partial_i \mu + \pi_j \partial_i V^j + \partial_j \left(\pi_i V^j\right).
\end{equation}
In particular, (\ref{eqn:pdefn}) gives 
\begin{equation}
  \partial_i P = \rho \partial_i \mu + \pi_j \partial_i V^j 
\end{equation}
so 
\begin{equation}
  -\partial_t \pi_j = \partial_i P + \partial_j (\pi_i V^j).
\end{equation}
With the identification 
\begin{equation}\label{eqn:stress}
  T_{ij} = \delta_{ij} P + \pi_i V_j,
\end{equation}
we ascertain that momentum is conserved and that 
\begin{equation}
  -\partial_t \pi_i = \partial_j T_i^j.
\end{equation}
(\ref{eqn:stress}) represents the ideal hydrodynamic expression for the stress tensor, in agreement with \cite{lt4dbreakdown_Lucas}.  The charge current $J_i = \rho V_i$, but as in general fracton hydrodynamics \cite{fracton_hydro}, it is not appropriate to think of the current as a vector; instead, one should think of currents in alternative representations of the spatial symmetry group. The practical consequence of this is that $V_i$ will (as we see) usually itself be a total derivative.  

\subsection{Enumerating conserved charges}
For any integrable function $f(x_1,x_2,\dots, x_d)$, we define a ``multipolar" charge associated to $f$ as 
\begin{equation}
  Q_f := \int \mathrm{d}^d x f(x_1,x_2,\dots,x_d)\rho.
\end{equation}
We say that $Q_f$ is conserved if \begin{equation}
    \frac{\mathrm{d} Q_f}{\mathrm{d}t} := \lbrace Q_f , H \rbrace= 0 .
\end{equation} 
If $F$ and $G$ are conserved by the dynamics, it follows from the Jacobi identity that $\lbrace F,G \rbrace $ is also conserved.   This places strong constraints on the kinds of fracton fluids that are allowed.

Define total momentum 
\begin{equation}
  \Pi_i  = \int \mathrm d^d x\, \pi_i .
\end{equation}
Recalling the definition made in (\ref{eqn:pb}), we immediately see that 
\begin{equation}\label{eqn:partial_conserv}
  \left\lbrace Q_f ,  \Pi_i \right\} = \int \mathrm{d}^d x \frac{\partial f}{\partial x_i} \rho = Q_{\partial_i f}.
\end{equation}
That is, provided that momentum is conserved by dynamics and a given $Q_f$ is conserved, every charge corresponding to any number of partial derivatives of $f$ is also conserved. 
We consider a number of examples where we demonstrate that a particular $Q_f$ is conserved; based on the discussion above to provide the fact that derivatives of $f$ also generate conserved charges, and do not need to be listed explicitly. In particular, the conservation of a charge generated by a polynomial of finite degree in $x_i$ implies that the charges generated by all lower--degree polynomials are also conserved. 

Now consider the transformation 
\begin{equation}
  v_i \rightarrow v_i + \partial_i f \label{eq:vtrans}
\end{equation}
We see, under this transformation, $H$ transforms as 
\begin{equation}
  \delta H =  \int \mathrm d ^d x \,  \frac{\delta H}{\delta v_i}\partial_if= - \int \mathrm d ^d x \, f \partial_i \frac{\delta H}{\delta v_i} = -\{H, Q_f\}.
\end{equation}
Evidently, the charge generated by $f$ is conserved if and only if $H$ is invariant under the transformation (\ref{eq:vtrans}).
We regard this fact as a consequence of the multipole algebra \cite{multipole_gromov}.

Pragmatically, this will allow us to construct kinetic terms which preserve charges generated by arbitrary $f$. We can also see why $v_i$ is more natural than $\pi_i$, since the ``shift symmetry" demanded by $Q_f$ is realized in a simpler way in (\ref{eq:vtrans}).
In particular, if $D_i$ is a collection of $d$ differential operators (in $d$ spatial dimensions) obeying
\begin{equation}
  D_i\partial^i f = 0,
\end{equation}
whenever $Q_f$ is conserved, then
\begin{equation}\label{eqn:kintrm}
  T = \frac{1}{2} (D_iv^i)^2 
\end{equation}
is a kinetic term which preserves the charge generated by $f$. In general it appears to be possible to ``eyeball" the sensible choices of $D_i$ which involve the fewest derivatives.

\subsection{Multipole conservation}
We wish to construct a family of models for each integer $n$ that conserve the first $n$ multipole moments. 
Pursuant to this, 
choose some fixed polynomial $f_n$ of degree $n$ and define 
\begin{equation}
  Q_{f_n} = \int \mathrm{d}^d x \, f_n  \rho. 
\end{equation}
for each positive integer $n$. 
Explicitly, we desire  
to construct a kinetic motif which preserves all possible choices of $Q_{f_n}$ for fixed $n$ (and all $m < n$). 
Pursuant to this, we define 
\begin{equation}\label{eqn:operator_kernel}
    D_{i_1,i_2,\dots,i_n} = \partial_{i_1} \partial_{i_2} \cdots \partial_{i_n}.
\end{equation}
Since any polynomial of degree less than $n$ is in the kernel of (\ref{eqn:operator_kernel}), we immediately see that 
\begin{equation}\label{eqn:Hdefn}
  \mathcal H = \frac{\rho}{2} \left( D_{i_1,i_2,\dots,i_n} v_{i_{n+1}} \right)\left(D^{i_1,i_2,\dots,i_n} v^{i_{n+1}}\right) + \frac{1}{2}\rho^2
\end{equation}
preserves $Q_{f_n}$.\footnote{Note that we have not chosen the only way to contract indices above together.  Enumerating the additional allowed tensors is straightforward (but tedious) and we will not do it here. See e.g. \cite{fracton_hydro,conservation_Nandkishore,hart2021hidden}.}
Without loss of generality, we are setting some prefactors to unity to simplify the resulting equations.  This system is in equilibrium when $\rho$ is constant and $\pi_i$ any polynomial of degree less than $n$.
We will now examine the properties of this system near equilibrium with and without the breaking of space--inversion and time--reversal symmetries. 

We start with a system with both inversion and time-reversal symmetry.
While it is possible to compute an explicit expression for the stress tensor of this system, we elect to compute it only to linear order in perturbations from equilibrium. We denote the first order deviation of some variable from its equilibrium value with a $\delta$. 
From (\ref{eqn:Vdefn}) and (\ref{eqn:Hdefn}), one can see that each $V_i$ vanishes in equilibrium and therefore that only the pressure is nontrivial at first order.
Once again appealing to (\ref{eqn:Vdefn}), one discovers that 
\begin{subequations}
  \begin{align}
    \partial_t \delta \rho &= - (-1)^n (\partial_i \partial^i)^{n} \partial_j \delta \pi^j \\
    \partial_t \delta \pi_i &= -\partial_i \delta \rho
  \end{align}
\end{subequations}
and, in turn, $\omega = \pm |k|^{n +1}$.  Hence we find ``magnon-like" modes but with arbitrary weak dispersion relations.

A curious feature of these systems is that the momentum susceptibility \begin{equation}
    \chi_{PP} \sim \frac{\pi_i}{V_i} \sim k^{-2n}.
\end{equation}
This generalizes the result of \cite{lt4dbreakdown_Lucas} to general multipole-conserving models with $n>1$.

\subsection{Breaking time-reversal symmetry in one dimensional models}

For the sake of convenience, we now restrict ourselves to a single spatial dimension.  We examine the properties of the multipole conserving Hamiltonians (\ref{eqn:Hdefn}) under the breaking of space--inversion and time--reversal symmetry. 
Consider 
\begin{equation}
 \mathcal{H} = \mathcal{H}_{\mathrm{T}} +  \mathcal U, \label{eq:HTU}
\end{equation}
where $\mathcal{H}_{\mathrm{T}}$ is a time-reversal symmetric Hamiltonian and \begin{equation}
    \mathcal U = \gamma \partial_x^m \rho \partial_x^n v.
\end{equation}  
Let us note the symmetries of $\mathcal{U}$.  It is always time-reversal odd, because under time-reversal (T): \begin{subequations}\label{eq:Ttrans}
  \begin{align}
      \mathrm{T}\cdot \rho &= \rho, \\
      \mathrm{T}\cdot v &= -v.
  \end{align}
\end{subequations}
Since under spatial inversion (or parity, P, in one dimension): 
\begin{subequations}\label{eq:Ptrans}
  \begin{align}
      \mathrm{P}\cdot \rho &= \rho, \\
      \mathrm{P}\cdot v &= -v,
  \end{align}
\end{subequations}
we conclude that $\mathcal{U}$ is parity-odd/even whenever $n+m$ is even/odd.  These facts will prove useful below, because we will want to consider the behavior of our theories when we either break time-reversal alone, or time-reversal along with parity.

The equations of motion for (\ref{eq:HTU}) are:
\begin{equation}\label{eqn:broken_eom}
  \begin{split}
    \partial_t v &= - \partial_x \left( \frac{1}{2}(\partial_x^n v)^2 + \rho +\gamma (-1)^m \partial_x^{n + m} v   \right) \\
    \partial_t \rho &= - \partial_x^{n+1} (-1)^n\left( \rho \partial_x^n v + \gamma \partial_x^m \rho\right).
  \end{split}
\end{equation}
Expanding in perturbations from equilibrium, Fourier transforming, and solving the resulting relation between frequency and wave number, we find 
\begin{equation}\label{eqn:disp_cont_broken}
  \omega = k^{n+1} \left[ \gamma \lambda(n,m) k^m \pm \sqrt{1  - \gamma^2\lambda(n+1,m)^2  k^{2m}  }\right]
\end{equation}
with 
\begin{equation}
  \lambda(n,m) =  \cos\left[\frac{\pi}{2}(n - m)\right] \cos[\pi(n+m)].
\end{equation}
We will continue to use this definition of $\lambda(n,m)$ for the remainder of this paper. 
The key features of $\lambda(n,m)$ are that  $\lambda(n,m)$ is only nonzero when $n$ and $m$ are of the same parity in $\mathbb Z_2$ and in this case, $|\lambda(n,m)| = 1$. 
In some sense, the symmetry  properties of the system at hand are captured entirely by  $\lambda$ and we find that precisely this function appears in other examples.
\begin{table}
  \centering
  
  \begin{tabular}{|c|c|}
    \hline
    $\mathrm P \,\,\mathcal U = \mathcal U$  & $\mathrm P \,\,\mathcal U = -\mathcal U$ \\
    \hline
    $n+m$ odd &\ $n+m$ even\\ \hline
    $\omega = \pm k^{n+1} \left(1 - \gamma^2 k^{2m} \right)$ &  $\omega =  k^{n+1}( \pm 1 + \sigma \gamma k^m)$\\
    \hline
  \end{tabular}
  \caption{\label{tab:small_k} Dispersion relation in linear response given by (\ref{eqn:broken_eom}) at small  $k$ for broken T and broken/unbroken P. $\sigma$ is a fixed real number determined by $n$ and $m$ so that $\sigma^2 = 1$.}
\end{table}

Small $k$ expansions of (\ref{eqn:disp_cont_broken}) are available in Table \ref{tab:small_k}. 
We are primarily interested in $m = 0$ and $m=1$ special cases of $\mathcal{U}$ because inflating $m$ provides corrections to the dispersion relation that 
are further and further from leading order; however, depending on whether we want parity to be broken or not, we must consider both the cases $m=0$ and $m=1$. 

Observe that when parity is preserved, the dispersion relations do not qualitatively change.  Indeed, from Table \ref{tab:small_k}, the dispersion relation modified only by O(1) constants, or at subleading orders in derivatives -- either of these effects however could also arise from time reversal symmetric perturbations.  However, when parity is broken, there is always a ``drift" term which is an odd integer power, such as $\gamma k^{2\ell+1}$; this is intuitive, since $\omega(\gamma,k) =\omega(-\gamma,-k)$.

It is instructive to consider particular instances of the above construction. First, we consider the dipole conserving case with $m=1$,
\begin{equation}\label{eq:dip}
  \mathcal H = \frac{\rho}{2}(\partial_x v)^2 + \frac{1}{2}\rho^2 + \gamma \partial_x \rho \partial_x v.
\end{equation}
Notice that the symmetry--breaking term, $\partial_x\rho \partial_x v$, is odd under space inversion and odd under time--reversal symmetry. Referring to (\ref{eqn:disp_cont_broken}), we see that the leading order dispersion relation acquires a sub--leading order drift term 
\begin{equation}
  \omega = \pm k^{2} - \gamma k^{3}.
\end{equation}
The choice $m=1$ seems more interesting than $m = 0$ because the latter gives 
\begin{equation}
  \omega = \pm k^{2}(1 - \gamma^2)
\end{equation}
which is qualitatively similar to the unbroken case. 
By contrast, in the quadrupole conserving case with $m = 0$,
\begin{equation}
  \mathcal H = \frac{\rho}{2}(\partial_x^2 v)^2 + \frac{1}{2}\rho^2 + \gamma \rho \partial_x^2 v
\end{equation}
leads to 
\begin{equation}\label{eq:qipd}
  \omega = \pm k^3 + \gamma k^3.
\end{equation}
In this case, the leading order dispersion relation is already modified by the ``drift" term proportional to $\gamma$.

\subsection{(Generalized) subsystem symmetries}

We now turn our attention to a family of models which exhibit a so--called ``subsystem symmetry" \cite{IaconisVijayNandkishore} (along with some multipolar generalizations thereof). We wish to construct a Hamiltonian density which preserves some number of multipole moments on every $d-1$ dimensional subset with a single fixed coordinate. In two spatial dimensions, this would mean that charge, dipole moment, etc are fixed on every line of fixed $x$ and every line of fixed $y$. 
Explicitly, in this setting, we demand
\begin{equation}
  \frac{\mathrm d}{\mathrm d t} \int_{x = a}\mathrm{d}y \,y^\alpha \rho  = \frac{\mathrm d}{\mathrm d t}\int_{y = b} \mathrm{d}x\, x^\beta \rho  =  0
\end{equation}
for integers $\alpha$ and $\beta$  less than or equal to some fixed positive integers $n_1$ and $n_2$ respectively. 
Recalling the discussion above, this amounts to conserving any charges of the form 
\begin{equation}\label{eqn:multipCharg}
  f(x,y) = \sum_{m_1=0}^{n_1} y^{m_1}f_1(x) + \sum_{m_2=0}^{n_2} x^{m_2}f_2(y) .
\end{equation}
Keeping previous constructions in mind, we seek to manufacture a kinetic term invariant under (\ref{eq:vtrans}).
Defining 
\begin{equation}
    K := \partial_x v_y + \partial_y v_x,
\end{equation}
and 
\begin{equation}
    D_2[v]:= \partial_x^{n_1} \partial_y^{n_2} K, 
\end{equation}
we see that $D_2[v]$ is sufficient for our purposes because
\begin{equation}
    \partial_x^{n_1 + 1} \partial_y^{n_2+1} f(x,y) = 0. 
\end{equation}
Generalizing this to an arbitrary number of spatial dimensions is not complicated. 
The following Hamiltonian includes both time-reversal breaking ($\gamma$) and the appropriate kinetic motifs to enforce the generalized subsystem symmetry above:
\begin{equation}\label{eqn:subsysham}
  \mathcal H = \frac{\rho}{4} D_2[v]^2 + \frac{1}{2}\rho^2  + \frac{\gamma}{2} \rho D_2[v].
\end{equation}
Using the equations of motion given in (\ref{eqn:eomsfr}), we find that within linear response:
\begin{equation}
  \begin{split}
    \partial_t \rho &=  (-1)^{n_1 + n_2} \partial_x^{n_1+1}\partial_y^{n_2 + 1} (\rho D_2[v] + \gamma \rho ) \\ 
    \partial_t v_i &= -\partial_i \left( \rho + \frac{\gamma}{2} D_2[v]\right). 
  \end{split}
\end{equation}
In order to obtain dispersion relations for this system, we must only consider that $\mathrm K$ vanishes in equilibrium and thus dispersions are given by 
\begin{equation}
  - \ii \omega
  \begin{pmatrix}
    \delta \rho_{\bm{k}} \\
    \delta \pi_{x,\bm{k}} \\
    \delta \pi_{y,\bm{k}}
  \end{pmatrix} = 
  \begin{pmatrix}
    -\gamma (-\ii)^{n_1 + n_2}  k_x^{n_1+1}  k_y^{n_2 + 1} & - \ii k_x^{2n_1 + 1} k_y^{2n_2 + 2} &  -\ii k_x^{2n_1 + 2}  k_y^{2n_2 +1} \\
    - \ii k_x &  \ii^{n_1 + n_2} \frac{\gamma}{2}  k_x^{n_1 + 1}  k_y^{n_2 + 1} &  \ii^{n_1 + n_2}\frac{\gamma}{2} k_x^{n_1 + 2}  k_y^{n_2} \\ 
    -\ii k_y & \ii^{n_1+n_2} \frac{\gamma}{2}  k_x^{n_1 } k_y^{n_2 + 2} & \ii^{n_1+n_2}\frac{\gamma}{2} k_x^{n_1+1}k_y^{n_2+1}
  \end{pmatrix}
  \begin{pmatrix}
    \delta \rho_{\bm{k}} \\
    \delta \pi_{x,\bm{k}} \\
    \delta \pi_{y,\bm{k}}
  \end{pmatrix} .
\end{equation}
This gives 
\begin{equation}\label{eqn:subsysdisp}
  \omega =  k_x^{n_1 + 1}k_y^{n_2 + 1} \left(    
    \gamma \lambda(n_1+n_2+1,0) \pm  \sqrt{2 - \gamma^2\lambda(n_1+n_2,0)^2 }
  \right).
\end{equation}
This is very similar in form to the  result given in (\ref{eqn:disp_cont_broken}). 
A notable difference is that, since $\mathrm K$ is even under space--inversion and odd under time--reversal, inversion is broken when $n_1+n_2$ is odd, or when $n_1+n_2 + 1$ is even, which is the circumstance wherein we see a drift term appear in (\ref{eqn:subsysdisp}).

 The careful reader may notice that we could have defined a more general kinetic term $K$:
\begin{equation}\label{eq:kgen}
  K:= c_1 \partial_x v_y + c_2 \partial_y v_x 
\end{equation}
without disrupting the symmetries of (\ref{eqn:subsysham}). 
Indeed, one may wonder if additional symmetries could be achieved by particular choices of $c_1$ and $c_2$. 
Since partial derivatives commute, under (\ref{eq:vtrans}), 
\begin{equation}
  K \rightarrow K + (c_1 + c_2) \partial_x \partial_y f
\end{equation}
for generic $f$. Of course, if $c_1+c_2=0$, then we effectively require 
\begin{equation}\label{eq:accidental}
  \frac{\mathrm d}{\mathrm d t}Q_f = 0
\end{equation}
for \textit{every} function $f$. This is only possible if $\rho$ is static. This is to say that the choice of $c_1$ and $c_2$ is unimportant qualitatively except in the case given by (\ref{eq:accidental}), in which case there is almost no dynamics.  This is slightly curious, since this choice corresponds to a motif where $K$ is the analogue of ``vorticity" in a regular fluid.

Another argument that one cannot simply add a ``vorticity" kinetic term to $\mathcal{H}$ is as follows. Using (\ref{eq:kgen}), 
\begin{equation}
    \partial_i \frac{\delta }{\delta v_i} \rho D_2[v_i]^2 = 2(-1)^{n_1 + n_2}(c_1+c_2 )\partial_x^{n_1+1} \partial_y^{n_2 + 1} \left(\rho D_2[v_i] \right).
\end{equation}
Of course, if $c_1 + c_2 = 0$, then this means that $\partial_t \rho = 0$.
It is straightforward to verify then that $K$ is static and therefore that $v_i$ is static provided that $\partial_i \rho$ also vanishes. 
In fact, in this particular setting, any function $\rho(x)$ may be chosen as the equilibrium value of $\rho$.  Then, we find that \begin{equation}
    \partial_t v_i = -\partial_i \rho,
\end{equation}
which implies that the velocity field is in general a linear function of $t$, since the right hand side is independent of $t$.

\section{Towards dissipative hydrodynamics}\label{towards_diss}
In this section, we will provide a few brief comments about the dissipative corrections to ideal hydrodynamics.  These are quite interesting, because following \cite{lt4dbreakdown_Lucas} they are expected to lead to new non-equilibrium universality classes which generalize the Kardar-Parisi-Zhang fixed point \cite{kpz,PhysRevE.90.012124}.  However, an exhaustive analysis of these effects is beyond this paper: in particular, because with the breaking of spacetime symmetries, we do not yet have a complete understanding of the allowed dissipative coefficients within hydrodynamics. Nevertheless, we will present some preliminary thoughts about what we expect, and hope to address these questions more systematically in the near future.
\subsection{Momentum relaxation}
One way to predict dissipative corrections to hydrodynamics is to relax momentum in a self-consistent way.  When this is done, we expect to reproduce the subdiffusive theories of \cite{fracton_hydro}, albeit now with the possibility of including inversion-breaking terms as well. 

Let us restrict our attention only to models in a single space dimension for the sake of convenience. 
Suppose we replace the equations of motion given in (\ref{eqn:eomsfr}) with 
\begin{equation}\label{eqn:eomsdisp}
  \begin{split}
    \partial_t \rho = -\partial_i \frac{\delta H}{\delta v_i} \\
    \partial_t v_i = -\partial_i \frac{\delta H}{\delta \rho} + (\partial_i v_j - \partial_j v_i)\frac{\delta H}{\delta v_j}- \beta v_i.
  \end{split}
\end{equation}
Note that $\beta$ is the relaxation rate for momentum density.  The dispersion relation of this system is given by 
\begin{equation}
  \det
  \begin{pmatrix}
    \gamma (-1)^{n + 1} (\ii k)^{n + m + 1} + \ii \omega & (-1)^{n +1 } (\ii k)^{2n + 1} \\
    -\ii k & \gamma (-1)^{m+1} (\ii k)^{n + m + 1} + \ii \omega - \beta
  \end{pmatrix}
  =0
\end{equation}
or equivalently
\begin{equation}
  \omega = -\ii \beta   + \gamma\lambda(n,m) k^{m + n + 1}  \pm \sqrt{  k^{2n + 2} - \left[k^{n + m + 1} \gamma \lambda(n+1,m)  + \beta  \right]^2 }.
\end{equation}
As in conventional hydrodynamics in the presence of momentum relaxation \cite{Grozdanov:2018fic,Baggioli:2019jcm}, there are two modes that exist, with dispersion relations as $k\rightarrow 0$: \begin{subequations}
\begin{align}
    \omega &\approx -\mathrm{i}\beta, \\
    \omega &\approx  \gamma \lambda(n,m) k^{m+n+1}-\frac{\mathrm{i} k^{2n+2}}{2\beta}.
\end{align}
\end{subequations}
The former corresponds to the finite relaxation rate for momentum density, while the latter corresponds to a subdiffusive mode for charge, with possible drift in the presence of inversion-breaking.
We see that, again, the presence of a leading order drift term depends on whether or not $n + m$ is even, as in Table \ref{tab:small_k}.   Note that the power of the drift term will generically be either $k^{n+1}$ or $k^{n+2}$ depending on whether $n$ is even or odd.  

This is actually somewhat non-trivial: as was emphasized already in \cite{fracton_hydro}, in general it is not the case that one can write down the lowest order coefficients in the higher-rank currents $J_{i_1\cdots i_{n+1}}$ in a fracton fluid.  Nevertheless, the model above implies that one must be able to write down these leading order terms for dissipationless drift in the presence of inversion breaking.  In particular, we predict that for a dipole-conserving model, one can only write down $J_{xx}= \partial_x \rho$ when inversion and time-reversal are broken, yet when quadrupoles are conserved, we can write down $J_{xxx} = \rho$!



\subsection{Instabilities}
In this subsection, we will predict the upper critical dimension below which hydrodynamics is unstable to fluctuations. We again assume momentum is conserved. We will focus on multipole-conserving theories in the discussion for simplicity, though similar power counting should hold for other models.  Following \cite{fracton_hydro} and the discussion above, we predict that the dynamical critical exponent of dissipation is \begin{equation}
    z = 2n+2. \label{eq:zdef}
\end{equation}
This is important, as this will fix the relative scaling of time and space in our power-counting arguments.  The reason for this is that, as in the ordinary KPZ analysis, one wishes to study the breakdown of hydrodynamics at the propagating wavefront of an excitation, which in a multipole-conserving theory, will have a dissipationless part $\omega \sim k^{n+1}+\cdots$.  

The equations of motion are
   \begin{subequations}
   \begin{align}
   \partial_t \pi_i  + \partial_j T^{j}_i &= 0, \\
   \partial_t \rho + \partial_{i_1} \partial_{i_2}\dots\partial_{i_{n+1}} J^{i_1 i_2 \dots i_{n+1}} &= 0. 
   \end{align}
   \end{subequations} 
We add noise in the form of $\tau_{ij}$  and $\xi_{i_1 i_2\dots i_{n+1}}$ so that the dynamics with fluctuations are related to those without by 
        \begin{subequations}
            \begin{align}
                T_{ij} &\rightarrow T_{ij} + \tau_{ij} \\
                J_{i_1 i_2\dots i_{n+1}} &\rightarrow J_{i_1 i_2\dots i_{n+1}} + \xi_{i_1 i_2\dots i_{n+1}}
            \end{align}
        \end{subequations}
Above, $\tau_{ij}$ and $\xi_{i_1\dots}$ are Gaussian white noise, with variances given by 
    \begin{subequations} \label{eq:gaussian}
    \begin{align}
    \langle \tau_{ij} \tau_{lm} \rangle &= 2\eta_{ijlm} \delta(t) \delta^d(x), \\
    \langle \xi_{i_1i_2\dots i_{n+1}} \xi_{j_1 j_2 \dots j_{n+1}} \rangle &= C_{i_1\dotsi_{n+1} j_1 \dots j_{n+1}} \delta(t) \delta^d(x) .
    \end{align}
    \end{subequations}
$\eta$ and $C$ here represent tensors proportional to dissipative coefficients within hydrodynamics.  In fluctuating hydrodynamics, we must take these noise terms to be marginal.  Combining (\ref{eq:zdef}) and (\ref{eq:gaussian}), we find that $\tau_{ij} \sim \xi_{i_1\cdots i_{n+1}} \sim k^{n+1+d/2}$.   (By this power counting, note that $\omega \sim k^{2n+2}$.)  But since the dimensions of currents and densities must be related, we can deduce that $\rho \sim k^{-2n-2 + (n+1)} \xi \sim k^{d/2}$, while $\pi_i \sim k^{d/2-n}$.

The leading order nonlinearity arises in the pressure $P(\rho)$, and so when expanding the equations about equilibrium ($\delta \rho = \rho - \rho_{\mathrm{eq}}$):
    \begin{subequations}
    \begin{align}
        \partial_t \delta \pi_i + \frac{1}{\chi}\partial_x \delta \rho + \lambda \delta \rho \partial_x \delta \rho + \lambda' \delta \rho^2 \partial_x \delta \rho + \partial_x \tau^{xx} +\dots =0, \\
        \partial_t \delta \rho - A \partial_x^{2n+1} v_x + \frac{C}{\chi}\partial_x^{2n+2} \delta \rho + \partial_x^{n+1} \xi_{xx\dots x}+\dots = 0  ,
    \end{align}
    \end{subequations}
we find that the dimension of the coefficient $\lambda \sim k^{n+2+d/2}k^{-1-d} \sim k^{n+1 - d/2}$. Thus this is a relevant perturbation whenever \begin{equation}
    d<2n+2.
\end{equation}
This means that $2n+2$ is the upper critical dimension of the momentum-conserving theory with $n$-pole conservation.  For large $n$, this can be arbitrarily large.  Below this upper critical dimension, we expect that this theory will flow towards a multipolar generalization of the KPZ fixed point, as was explained in detail for the case $n=1$ in \cite{lt4dbreakdown_Lucas}.

\section{Microscopic models in 1d}\label{microscopic}

In this section, we present a list of microscopic Hamiltonian models which exhibit both momentum and multipole conservation in one spatial dimension.  We will present these systems as Hamiltonian dynamical systems, and so strictly speaking all of these models also have energy conservation.  Following \cite{lt4dbreakdown_Lucas}, it is possible to relax energy conservation by adding suitable noise; one can also simply make some coefficients time-dependent if desired.  These constructions may be useful in actually carrying out large scale numerical simulations, in order to look for the non-equilibrium fixed points predicted above.  Unfortunately, due to the extremely slow relaxation predicted above, it will be quite challenging to run the simulation for long enough to detect the new physics!

\subsection{Constructing the Hamiltonian}

Consider $N$ particles, arranged on a one dimensional line, labeled by $i=1,\ldots, N$.  Their position and momentum are given by the canonically related $x_i$  and $p_i$: \begin{equation}
    \lbrace x_i, p_j\rbrace = \delta_{ij}.
\end{equation} In this language, we define the multipole moments 
\begin{equation}
  Q_n = \sum_{i = 1}^N x_i^{n}
\end{equation}
so that $Q_0$ is total system charge, $Q_1$ is the total dipole moment and so on. We also define 
\begin{equation}
  P = \sum_{i =1}^N p_i
\end{equation}
to be the total momentum of the system. We aim to construct a family of models which leave the first few multipole moments and total momentum invariant under time evolution. 
Namely, we are looking for a Hamiltonian $H$ such that \begin{equation}
    \lbrace H,P\rbrace = \lbrace H,Q_0\rbrace = \cdots = \lbrace H,Q_n\rbrace = 0. \label{eq:sec4cons}
\end{equation}
Note that 
\begin{equation}
 \lbrace H,Q_n\rbrace = \left\{\sum_{i=1}^N x^n_i,H\right\}  = n\sum_{i=1}^N   x^{n-1}_i \frac{\partial H }{\partial p_i} = 0.
\end{equation}

To help construct such an $H$, for positive integers $n$ and $m$, let us define 
\begin{equation}
  \Lambda_{n,i} = 
  \begin{pmatrix}
    p_i & p_{i+1} & \dots & p_{i + n } \\
    1   &    1    & \dots & 1 \\
    x_i & x_{i+1} & \dots & x_{i + n } \\ 
    x_i^2 & x_{i+1}^2 & \dots & x_{i+n}^2 \\
    \vdots &      &   &  \vdots \\ 
    x_i^{n-1} & x_{i+1}^{n-1} & \dots & x_{i +n}^{n-1}
  \end{pmatrix}
\end{equation}
and \begin{equation}
    L_{n,i} = \det \Lambda_{n,i}. 
\end{equation}
Since $L_{n,i}$  is a sum of terms linear in momenta, 
\begin{equation}\label{eqn:npolcon}
  \sum_{j = 1}^N x_j^{n-1} \frac{\partial L_{n,i}}{\partial p_j} = 
  \det 
  \begin{pmatrix}
    x_i^{n-1} & x_{i+1}^{n-1} & \dots & x_{i +n}^{n-1} \\ 
    1   &    1    & \dots & 1 \\
    x_i & x_{i+1} & \dots & x_{i + n } \\ 
    x_i^2 & x_{i+1}^2 & \dots & x_{i+n}^2 \\
    \vdots &      &   &  \vdots \\ 
    x_i^{n-1} & x_{i+1}^{n-1} & \dots & x_{i +n}^{n-1}
  \end{pmatrix} = 0
\end{equation}
by linear dependence.  In fact, we conclude that \begin{equation}
    \lbrace Q_m , L_{n,i} \rbrace = 0, \;\;\; (m\le n).
\end{equation}
As a consequence, $L_{n,i}$ is an invariant which we can use to start building invariant Hamiltonians.

It is instructive to simplify the form of $L_{n,i}$ somewhat.  Assuming that $i\le j \le i+n$, the coefficient of $p_j$ in $L_{n,i} $ is given by 
\begin{equation}\label{eqn:pcon}
  \frac{\partial L_{n,i}}{\partial p_j} = (-1)^{j - i}\det
  \begin{pmatrix}
    1 & 1 & \dots & 1 & 1& \dots & 1  \\
    x_i & x_{i+1} & \dots & x_{j - 1} & x_{j +1} & \dots & x_{i + n} \\
    x_i^2 & x^2_{i+1} & \dots & x^2_{j - 1} & x^2_{j +1} & \dots & x^2_{i + n} \\
    \vdots & &&&&& \vdots \\
    x_i^{n-1} & x_{i+1}^{n-1} & \dots & x^{n-1}_{j - 1} & x^{n-1}_{j +1} & \dots & x^{n-1}_{i + n} \\
  \end{pmatrix}. 
\end{equation}
Noting that 
\begin{equation}
  a^m - b^m = (a - b) \sum_{l = 1}^m a^{m - l}b^{l-1},
\end{equation}
and subtracting one column appearing in (\ref{eqn:pcon}) from another, we see that for any pair of integers $(a,b)$ so that  $a \neq j \neq b$ and 
$a \neq b$ with both  $i \leq a,b \leq i + n$, $x_a - x_b$ is a factor of $\frac{\partial L_{n,i}}{\partial p_j}$. 
Realizing that $L _{n,i}$ is linear in each momenta, that the coefficient of each term in this multinomial (in $x_k$s) coefficient is $\pm 1$, and by power counting, we deduce that, up to an overall sign,
\begin{equation}
  L_{n,i} = \pm \sum_{j = i}^{i + n} p_j \prod_{\substack{i \leq u \leq v \leq i + n \\ u,v \neq j}} (x_u - x_v).
\end{equation}
We immediately see that $L_{n,i}$ is invariant under $x_i \rightarrow x_i + c$ for all $i$, and therefore \begin{equation}
    \lbrace P,L_{n,i}\rbrace = 0.
\end{equation}
Thus $L_{n,i}$ can be used to write down a multipole-conserving kinetic motif.  Note that once $n>1$, $L_{n,i}$ is intrinsically nonlinear. 

Now, consider a Hamiltonian of the form
\begin{equation}\label{eqn:hmin}
  H = V(x_1,\ldots, x_N) + \sum_{i = 1}^{N-n} \frac{1}{2} L_{n,i}^2 .
\end{equation}
So long as $V$ is translation invariant, we are guaranteed that (\ref{eq:sec4cons}) is obeyed.  A minimal Hamiltonian corresponds to choosing \begin{equation}
  H = \sum_{i=1}^{N-1}\frac{1}{2}(1+x_i-x_{i+1})^2 + \sum_{i = 1}^{N-n} \frac{1}{2} L_{n,i}^2 .
\end{equation}
Note that we have chosen our potential energy such that equilibrium corresponds to (e.g.) $x_i=i$.
For any suitable choice of equilibrium, we must have that $L_{n,i}$ vanishes in equilibrium. One can see from (\ref{eqn:npolcon}) that $L_{n,i}$ vanishes at  $p_i = c$ for  all $i$.  

\subsection{Time-reversal breaking}
It is straightforward to incorporate broken time reversal symmetry (and/or broken parity).  Let $\Gamma_i$ be a function of only $x_k$s, which is invariant under translation and reasonably local.  Then we can try to add the following time-reversal breaking term to $H$: \begin{equation} \label{eqn:hbrok}
  \tilde H = \gamma \sum_{i=1}^{N-n}\Gamma_i L_{n,i}.
\end{equation}
Here $\gamma$ is a constant. 
In order to preserve a chosen equilibrium, we require that  
\begin{equation}
  \frac{\partial }{\partial x_m} \tilde H\big\rvert_{\mathrm{eq}} = \frac{\partial}{\partial p_m}\tilde H\big\rvert_{\mathrm{eq}} = 0
\end{equation}
for any integer $m$. 
Given that $L_{n,i}$ must vanish in equlibrium and that $\Gamma_i$ is assumed to be independent of momenta, these conditions are equivalent to 
\begin{equation}
\sum_{i = 1}^{N-n} \Gamma_i \frac{\partial L_{n,i}}{\partial p_m}\bigg\rvert_{eq} = \sum_{i =1 }^{N - n} \Gamma_i \frac{\partial L_{n,i}}{\partial x_m}\bigg\rvert_{eq} = 0.
\end{equation}
Again, the form of (\ref{eqn:npolcon}) guarantees that the second above inequality (namely conservation of momentum) holds and we are left only with the condition that
\begin{equation}\label{eqn:gammasym}
\sum_{i=1}^{N-n} \Gamma_i \frac{\partial L_{n,i}}{\partial p_m} \bigg\rvert_{eq} = 0 .
\end{equation}
Recalling the expression given in (\ref{eqn:pcon}), we find that 
\begin{equation}\label{eqn:lsim}
\frac{\partial L_{n,i}}{\partial p_m}\bigg\rvert_{x_j = j} =  (-1)^{m - i}\binom{n}{m - i}\prod_{j = 1}^{n-1} j!
\end{equation}
There are a large class of $\Gamma_i$ which may be chosen to satisfy (\ref{eqn:gammasym}) in equilibrium with (\ref{eqn:lsim}). In particular, we can take $\Gamma_i$ to be a polynomial of degree $<n$ in $x_i$ (or $x_{i+j}$ for some $j\ne 0$), because for any polynomial $A$ of degree less than $n$, for any $m$:
\begin{equation}
    \sum_{i = 0}^n (-1)^i \binom{n}{i} A(m+i) = 0. 
\end{equation}

\subsection{Quasinormal modes}
Now suppose we wish to conserve up to the $n$-pole moment. 
Without concerning ourselves over whether or not $H$ would be convergent, we consider the infinite chain limit ($N \rightarrow \infty$):
\begin{equation}
   H = \frac{1}{2} \sum_{i = -\infty}^\infty (1 + x_i - x_{i+1})^2 + L_i^2 + \gamma \Gamma_i L_i.
\end{equation}
Loosely, this is $H + \tilde H$ from (\ref{eqn:hmin}) and (\ref{eqn:hbrok}) respectively. 
There are a variety of suitable choices for $\Gamma_i$. As a minimal choice, we will require that $\Gamma_i$ 
be linear in positions and odd under space--inversion. Namely,
\begin{equation}
  \Gamma_i = x_i - x_{i-1}.
\end{equation}
However, the following results are insensitive to the details of $\Gamma_i$. Indeed,
assuming only that $\mathrm d \Gamma_i/\mathrm d x_j\rvert_{eq}$ is dependent only on $i-j$ is sufficient to ascertain dispersion relations up to constant coefficients. 
 Writing $x_k$ and $p_k$ to be the discrete Fourier transform of $x_n$ and $p_n$ respectively, and defining
 \begin{subequations}\label{eqn:SNdefn}
   \begin{align}
     \mathrm S &= (1 - \mathrm{e}^{-\ii k})^{n} \\ 
     \mathrm N &= \sum_{l = -\infty}^\infty \mathrm{e}^{\ii l k} \frac{\partial \Gamma_0}{\partial x_l} \\
     \mathrm M &= 4 \sin^2 \left(\frac{k}{2}\right)
   \end{align}
 \end{subequations}
 and we find that the equations of motion given by $H$ are 
 \begin{subequations}\label{eqn:2by2}
   \begin{align}
      \partial_t \delta x_k = c_n^2 \mathrm M ^{n} \delta p_k  + \gamma c_n \mathrm S^* \mathrm N^* \delta x_k  \\
     -\partial_t \delta p_k = \mathrm M \delta x_k + \gamma c_n \mathrm S \mathrm N \delta p_k
   \end{align}
 \end{subequations}
 with 
 \begin{equation}
   c_n = \prod_{m = 1}^{n-1} m! .
 \end{equation}
 For our particular choice of $\Gamma_i$,
  \begin{equation}
    \mathrm N = 1 - \mathrm{e}^{-\ii k}.
 \end{equation}
 Using, (\ref{eqn:2by2}), we can produce an exact dispersion relation:
\begin{equation}
    \omega =   c_n \gamma \text{Im}[\mathrm S \mathrm N]  \pm c_n \sqrt{ 4^{n+1} \sin^{2n+2}(k/2) - \gamma^2  \text{Re}[\mathrm S \mathrm N]^2}. \label{eq:omega43}
\end{equation}
 It follows immediately that $\omega$ is real valued for sufficiently small $\gamma$. 
 Momentum conservation guarantees that $\mathrm N$ vanishes at vanishing $k$. This fact immediately demonstrates that the leading $k$ behavior of $\omega$ cannot be reduced by breaking symmetry in this manner. 
 
 Expanding, now, in small $k$, we find that \begin{subequations}
   \begin{align}
       \mathrm{S} &\approx (\mathrm{i}k)^n + \frac{n}{2}k^{n+1}, \\
       \mathrm{N} &\approx \mathrm{i}k + \frac{k^2}{2}, \\
       \mathrm{M} &\approx k^2.
   \end{align}
 \end{subequations}
 Therefore, \begin{subequations}
   \begin{align}
       \mathrm{Im}(\mathrm{SN}) &\sim \left\lbrace \begin{array}{ll} k^{n+2} &\ n \text{ odd} \\ k^{n+1} &\ n \text{ even} \end{array}\right., \\
       \mathrm{Re}(\mathrm{SN}) &\sim \left\lbrace \begin{array}{ll} k^{n+1} &\ n \text{ odd} \\ k^{n+2} &\ n \text{ even} \end{array}\right..
   \end{align}
 \end{subequations}
 Observe that the propagating modes have, at leading order, $\omega\sim k^{n+1}$. When $n$ is even, we observe that the drift term in (\ref{eq:omega43}) must be subleading. These facts are precisely in agreement with our theory in Section \ref{ideal_hydro}.
  
 

 
\section{Conclusions}
To summarize, we have described infinitely many new universality classes of fracton hydrodynamics with both momentum conservation and multipolar or subsystem conservation laws.  We expect that all of these universality classes are -- in sufficiently low dimension -- unstable, similar to what was found in \cite{lt4dbreakdown_Lucas} for fluids with dipole and momentum conservation.  The models presented in the previous section can serve as concrete starting points for systematic numerical investigations of these new non-equilibrium fixed points in one-dimensional models;  however, we caution that due to the very large dynamical critical exponents expected for each new universality class, the time scales required to simulate the dynamics may be quite long (and thus require many computational resources).  

Beyond more direct investigations of dissipative dynamics in these new universality classes, which we expect will largely follow \cite{lt4dbreakdown_Lucas,paolotoappear}, we believe that it is particularly important to understand better the role of spacetime symmetries (such as time-reversal) in fracton hydrodynamics.  The models we constructed in this paper will provide a valuable starting point for any such future investigation, with or without momentum conservation.  

\section*{Acknowledgements} 
We thank Paolo Glorioso for helpful comments.  This work was supported by a Research Fellowship from the Alfred P. Sloan Foundation under Grant FG-2020-13795 (AL), by the Gordon and Betty Moore Foundation's EPiQS Initiative under Grant GBMF10279 (AL), and by the U.S. Air Force Office of Scientific Research under Grant FA9550-21-1-0195 (AO).  AL also acknowledges the hospitality of KITP, which is supported in part by the National Science Foundation under Grant No. NSF PHY-1748958.



\bibliography{refs.bib} 

\begin{thebibliography}{50}%
\makeatletter
\providecommand \@ifxundefined [1]{%
 \@ifx{#1\undefined}
}%
\providecommand \@ifnum [1]{%
 \ifnum #1\expandafter \@firstoftwo
 \else \expandafter \@secondoftwo
 \fi
}%
\providecommand \@ifx [1]{%
 \ifx #1\expandafter \@firstoftwo
 \else \expandafter \@secondoftwo
 \fi
}%
\providecommand \natexlab [1]{#1}%
\providecommand \enquote  [1]{``#1''}%
\providecommand \bibnamefont  [1]{#1}%
\providecommand \bibfnamefont [1]{#1}%
\providecommand \citenamefont [1]{#1}%
\providecommand \href@noop [0]{\@secondoftwo}%
\providecommand \href [0]{\begingroup \@sanitize@url \@href}%
\providecommand \@href[1]{\@@startlink{#1}\@@href}%
\providecommand \@@href[1]{\endgroup#1\@@endlink}%
\providecommand \@sanitize@url [0]{\catcode `\\12\catcode `\$12\catcode
  `\&12\catcode `\#12\catcode `\^12\catcode `\_12\catcode `\%12\relax}%
\providecommand \@@startlink[1]{}%
\providecommand \@@endlink[0]{}%
\providecommand \url  [0]{\begingroup\@sanitize@url \@url }%
\providecommand \@url [1]{\endgroup\@href {#1}{\urlprefix }}%
\providecommand \urlprefix  [0]{URL }%
\providecommand \Eprint [0]{\href }%
\providecommand \doibase [0]{http://dx.doi.org/}%
\providecommand \selectlanguage [0]{\@gobble}%
\providecommand \bibinfo  [0]{\@secondoftwo}%
\providecommand \bibfield  [0]{\@secondoftwo}%
\providecommand \translation [1]{[#1]}%
\providecommand \BibitemOpen [0]{}%
\providecommand \bibitemStop [0]{}%
\providecommand \bibitemNoStop [0]{.\EOS\space}%
\providecommand \EOS [0]{\spacefactor3000\relax}%
\providecommand \BibitemShut  [1]{\csname bibitem#1\endcsname}%
\let\auto@bib@innerbib\@empty
\bibitem [{\citenamefont {Prem}\ \emph
  {et~al.}(2018{\natexlab{a}})\citenamefont {Prem}, \citenamefont {Pretko},\
  and\ \citenamefont {Nandkishore}}]{emergent_Nandkishore}%
  \BibitemOpen
  \bibfield  {author} {\bibinfo {author} {\bibfnamefont {Abhinav}\ \bibnamefont
  {Prem}}, \bibinfo {author} {\bibfnamefont {Michael}\ \bibnamefont {Pretko}},
  \ and\ \bibinfo {author} {\bibfnamefont {Rahul~M.}\ \bibnamefont
  {Nandkishore}},\ }\bibfield  {title} {\enquote {\bibinfo {title} {Emergent
  phases of fractonic matter},}\ }\href {\doibase 10.1103/PhysRevB.97.085116}
  {\bibfield  {journal} {\bibinfo  {journal} {Phys. Rev. B}\ }\textbf {\bibinfo
  {volume} {97}},\ \bibinfo {pages} {085116} (\bibinfo {year}
  {2018}{\natexlab{a}})}\BibitemShut {NoStop}%
\bibitem [{\citenamefont {Prem}\ \emph {et~al.}(2019)\citenamefont {Prem},
  \citenamefont {Huang}, \citenamefont {Song},\ and\ \citenamefont
  {Hermele}}]{cage_Hermele}%
  \BibitemOpen
  \bibfield  {author} {\bibinfo {author} {\bibfnamefont {Abhinav}\ \bibnamefont
  {Prem}}, \bibinfo {author} {\bibfnamefont {Sheng-Jie}\ \bibnamefont {Huang}},
  \bibinfo {author} {\bibfnamefont {Hao}\ \bibnamefont {Song}}, \ and\ \bibinfo
  {author} {\bibfnamefont {Michael}\ \bibnamefont {Hermele}},\ }\bibfield
  {title} {\enquote {\bibinfo {title} {Cage-net fracton models},}\ }\href
  {\doibase 10.1103/PhysRevX.9.021010} {\bibfield  {journal} {\bibinfo
  {journal} {Phys. Rev. X}\ }\textbf {\bibinfo {volume} {9}},\ \bibinfo {pages}
  {021010} (\bibinfo {year} {2019})}\BibitemShut {NoStop}%
\bibitem [{\citenamefont {Prem}\ \emph
  {et~al.}(2018{\natexlab{b}})\citenamefont {Prem}, \citenamefont {Vijay},
  \citenamefont {Chou}, \citenamefont {Pretko},\ and\ \citenamefont
  {Nandkishore}}]{pinch_Nandkishore}%
  \BibitemOpen
  \bibfield  {author} {\bibinfo {author} {\bibfnamefont {Abhinav}\ \bibnamefont
  {Prem}}, \bibinfo {author} {\bibfnamefont {Sagar}\ \bibnamefont {Vijay}},
  \bibinfo {author} {\bibfnamefont {Yang-Zhi}\ \bibnamefont {Chou}}, \bibinfo
  {author} {\bibfnamefont {Michael}\ \bibnamefont {Pretko}}, \ and\ \bibinfo
  {author} {\bibfnamefont {Rahul}\ \bibnamefont {Nandkishore}},\ }\bibfield
  {title} {\enquote {\bibinfo {title} {Pinch point singularities of tensor spin
  liquids},}\ }\href {\doibase 10.1103/PhysRevB.98.165140} {\bibfield
  {journal} {\bibinfo  {journal} {Physical Review B}\ }\textbf {\bibinfo
  {volume} {98}},\ \bibinfo {pages} {165140} (\bibinfo {year}
  {2018}{\natexlab{b}})}\BibitemShut {NoStop}%
\bibitem [{\citenamefont {Slagle}\ \emph
  {et~al.}(2019{\natexlab{a}})\citenamefont {Slagle}, \citenamefont {Prem},\
  and\ \citenamefont {Pretko}}]{symmetric_Pretko}%
  \BibitemOpen
  \bibfield  {author} {\bibinfo {author} {\bibfnamefont {Kevin}\ \bibnamefont
  {Slagle}}, \bibinfo {author} {\bibfnamefont {Abhinav}\ \bibnamefont {Prem}},
  \ and\ \bibinfo {author} {\bibfnamefont {Michael}\ \bibnamefont {Pretko}},\
  }\bibfield  {title} {\enquote {\bibinfo {title} {Symmetric tensor gauge
  theories on curved spaces},}\ }\href {\doibase
  https://doi.org/10.1016/j.aop.2019.167910} {\bibfield  {journal} {\bibinfo
  {journal} {Annals of Physics}\ }\textbf {\bibinfo {volume} {410}},\ \bibinfo
  {pages} {167910} (\bibinfo {year} {2019}{\natexlab{a}})}\BibitemShut
  {NoStop}%
\bibitem [{\citenamefont {Song}\ \emph {et~al.}(2019)\citenamefont {Song},
  \citenamefont {Prem}, \citenamefont {Huang},\ and\ \citenamefont
  {Martin-Delgado}}]{twisted_Delgado}%
  \BibitemOpen
  \bibfield  {author} {\bibinfo {author} {\bibfnamefont {Hao}\ \bibnamefont
  {Song}}, \bibinfo {author} {\bibfnamefont {Abhinav}\ \bibnamefont {Prem}},
  \bibinfo {author} {\bibfnamefont {Sheng-Jie}\ \bibnamefont {Huang}}, \ and\
  \bibinfo {author} {\bibfnamefont {M.~A.}\ \bibnamefont {Martin-Delgado}},\
  }\bibfield  {title} {\enquote {\bibinfo {title} {Twisted fracton models in
  three dimensions},}\ }\href {\doibase 10.1103/PhysRevB.99.155118} {\bibfield
  {journal} {\bibinfo  {journal} {Phys. Rev. B}\ }\textbf {\bibinfo {volume}
  {99}},\ \bibinfo {pages} {155118} (\bibinfo {year} {2019})}\BibitemShut
  {NoStop}%
\bibitem [{\citenamefont {Slagle}\ and\ \citenamefont
  {Kim}(2017{\natexlab{a}})}]{fracton_Kim}%
  \BibitemOpen
  \bibfield  {author} {\bibinfo {author} {\bibfnamefont {Kevin}\ \bibnamefont
  {Slagle}}\ and\ \bibinfo {author} {\bibfnamefont {Yong~Baek}\ \bibnamefont
  {Kim}},\ }\bibfield  {title} {\enquote {\bibinfo {title} {Fracton topological
  order from nearest-neighbor two-spin interactions and dualities},}\ }\href
  {\doibase 10.1103/PhysRevB.96.165106} {\bibfield  {journal} {\bibinfo
  {journal} {Phys. Rev. B}\ }\textbf {\bibinfo {volume} {96}},\ \bibinfo
  {pages} {165106} (\bibinfo {year} {2017}{\natexlab{a}})}\BibitemShut
  {NoStop}%
\bibitem [{\citenamefont {Slagle}\ and\ \citenamefont
  {Kim}(2017{\natexlab{b}})}]{quantum_Kim}%
  \BibitemOpen
  \bibfield  {author} {\bibinfo {author} {\bibfnamefont {Kevin}\ \bibnamefont
  {Slagle}}\ and\ \bibinfo {author} {\bibfnamefont {Yong~Baek}\ \bibnamefont
  {Kim}},\ }\bibfield  {title} {\enquote {\bibinfo {title} {Quantum field
  theory of x-cube fracton topological order and robust degeneracy from
  geometry},}\ }\href {\doibase 10.1103/PhysRevB.96.195139} {\bibfield
  {journal} {\bibinfo  {journal} {Phys. Rev. B}\ }\textbf {\bibinfo {volume}
  {96}},\ \bibinfo {pages} {195139} (\bibinfo {year}
  {2017}{\natexlab{b}})}\BibitemShut {NoStop}%
\bibitem [{\citenamefont {Slagle}\ and\ \citenamefont {Kim}(2018)}]{xcube_Kim}%
  \BibitemOpen
  \bibfield  {author} {\bibinfo {author} {\bibfnamefont {Kevin}\ \bibnamefont
  {Slagle}}\ and\ \bibinfo {author} {\bibfnamefont {Yong~Baek}\ \bibnamefont
  {Kim}},\ }\bibfield  {title} {\enquote {\bibinfo {title} {X-cube model on
  generic lattices: Fracton phases and geometric order},}\ }\href {\doibase
  10.1103/physrevb.97.165106} {\bibfield  {journal} {\bibinfo  {journal}
  {Physical Review B}\ }\textbf {\bibinfo {volume} {97}} (\bibinfo {year}
  {2018}),\ 10.1103/physrevb.97.165106}\BibitemShut {NoStop}%
\bibitem [{\citenamefont {Slagle}\ \emph
  {et~al.}(2019{\natexlab{b}})\citenamefont {Slagle}, \citenamefont {Aasen},\
  and\ \citenamefont {Williamson}}]{foilated_Williamson}%
  \BibitemOpen
  \bibfield  {author} {\bibinfo {author} {\bibfnamefont {Kevin}\ \bibnamefont
  {Slagle}}, \bibinfo {author} {\bibfnamefont {David}\ \bibnamefont {Aasen}}, \
  and\ \bibinfo {author} {\bibfnamefont {Dominic}\ \bibnamefont {Williamson}},\
  }\bibfield  {title} {\enquote {\bibinfo {title} {Foliated field theory and
  string-membrane-net condensation picture of fracton order},}\ }\href
  {\doibase 10.21468/SciPostPhys.6.4.043} {\bibfield  {journal} {\bibinfo
  {journal} {SciPost Physics}\ }\textbf {\bibinfo {volume} {6}} (\bibinfo
  {year} {2019}{\natexlab{b}}),\ 10.21468/SciPostPhys.6.4.043}\BibitemShut
  {NoStop}%
\bibitem [{\citenamefont {Pretko}(2017)}]{higher_Pretko}%
  \BibitemOpen
  \bibfield  {author} {\bibinfo {author} {\bibfnamefont {Michael}\ \bibnamefont
  {Pretko}},\ }\bibfield  {title} {\enquote {\bibinfo {title} {Higher-spin
  witten effect and two-dimensional fracton phases},}\ }\href {\doibase
  10.1103/PhysRevB.96.125151} {\bibfield  {journal} {\bibinfo  {journal} {Phys.
  Rev. B}\ }\textbf {\bibinfo {volume} {96}},\ \bibinfo {pages} {125151}
  (\bibinfo {year} {2017})}\BibitemShut {NoStop}%
\bibitem [{\citenamefont {Devakul}\ and\ \citenamefont
  {Sondhi}(2017)}]{correlation_Sondhi}%
  \BibitemOpen
  \bibfield  {author} {\bibinfo {author} {\bibfnamefont {Trithep}\ \bibnamefont
  {Devakul}}\ and\ \bibinfo {author} {\bibfnamefont {S.}~\bibnamefont
  {Sondhi}},\ }\bibfield  {title} {\enquote {\bibinfo {title} {Correlation
  function diagnostics for type-i fracton phases},}\ }\href {\doibase
  10.1103/PhysRevB.97.041110} {\bibfield  {journal} {\bibinfo  {journal}
  {Physical Review B}\ }\textbf {\bibinfo {volume} {97}} (\bibinfo {year}
  {2017}),\ 10.1103/PhysRevB.97.041110}\BibitemShut {NoStop}%
\bibitem [{\citenamefont {Devakul}\ \emph {et~al.}(2019)\citenamefont
  {Devakul}, \citenamefont {You}, \citenamefont {Burnell},\ and\ \citenamefont
  {Sondhi}}]{fractal_Sondhi}%
  \BibitemOpen
  \bibfield  {author} {\bibinfo {author} {\bibfnamefont {Trithep}\ \bibnamefont
  {Devakul}}, \bibinfo {author} {\bibfnamefont {Yizhi}\ \bibnamefont {You}},
  \bibinfo {author} {\bibfnamefont {F.~J.}\ \bibnamefont {Burnell}}, \ and\
  \bibinfo {author} {\bibfnamefont {Shivaji}\ \bibnamefont {Sondhi}},\
  }\bibfield  {title} {\enquote {\bibinfo {title} {Fractal symmetric phases of
  matter},}\ }\href {\doibase 10.21468/scipostphys.6.1.007} {\bibfield
  {journal} {\bibinfo  {journal} {SciPost Physics}\ }\textbf {\bibinfo {volume}
  {6}} (\bibinfo {year} {2019}),\ 10.21468/scipostphys.6.1.007}\BibitemShut
  {NoStop}%
\bibitem [{\citenamefont {You}\ \emph {et~al.}(2018)\citenamefont {You},
  \citenamefont {Devakul}, \citenamefont {Burnell},\ and\ \citenamefont
  {Sondhi}}]{subsystem_You}%
  \BibitemOpen
  \bibfield  {author} {\bibinfo {author} {\bibfnamefont {Yizhi}\ \bibnamefont
  {You}}, \bibinfo {author} {\bibfnamefont {Trithep}\ \bibnamefont {Devakul}},
  \bibinfo {author} {\bibfnamefont {F.~J.}\ \bibnamefont {Burnell}}, \ and\
  \bibinfo {author} {\bibfnamefont {S.~L.}\ \bibnamefont {Sondhi}},\ }\bibfield
   {title} {\enquote {\bibinfo {title} {Subsystem symmetry protected
  topological order},}\ }\href {\doibase 10.1103/physrevb.98.035112} {\bibfield
   {journal} {\bibinfo  {journal} {Physical Review B}\ }\textbf {\bibinfo
  {volume} {98}} (\bibinfo {year} {2018}),\
  10.1103/physrevb.98.035112}\BibitemShut {NoStop}%
\bibitem [{\citenamefont {You}\ \emph {et~al.}(2020{\natexlab{a}})\citenamefont
  {You}, \citenamefont {Devakul}, \citenamefont {Burnell},\ and\ \citenamefont
  {Sondhi}}]{symmetric_You}%
  \BibitemOpen
  \bibfield  {author} {\bibinfo {author} {\bibfnamefont {Yizhi}\ \bibnamefont
  {You}}, \bibinfo {author} {\bibfnamefont {Trithep}\ \bibnamefont {Devakul}},
  \bibinfo {author} {\bibfnamefont {F.J.}\ \bibnamefont {Burnell}}, \ and\
  \bibinfo {author} {\bibfnamefont {S.L.}\ \bibnamefont {Sondhi}},\ }\bibfield
  {title} {\enquote {\bibinfo {title} {Symmetric fracton matter: Twisted and
  enriched},}\ }\href {\doibase 10.1016/j.aop.2020.168140} {\bibfield
  {journal} {\bibinfo  {journal} {Annals of Physics}\ }\textbf {\bibinfo
  {volume} {416}},\ \bibinfo {pages} {168140} (\bibinfo {year}
  {2020}{\natexlab{a}})}\BibitemShut {NoStop}%
\bibitem [{\citenamefont {You}\ \emph {et~al.}(2020{\natexlab{b}})\citenamefont
  {You}, \citenamefont {Devakul}, \citenamefont {Sondhi},\ and\ \citenamefont
  {Burnell}}]{fractonic_You}%
  \BibitemOpen
  \bibfield  {author} {\bibinfo {author} {\bibfnamefont {Yizhi}\ \bibnamefont
  {You}}, \bibinfo {author} {\bibfnamefont {Trithep}\ \bibnamefont {Devakul}},
  \bibinfo {author} {\bibfnamefont {S.~L.}\ \bibnamefont {Sondhi}}, \ and\
  \bibinfo {author} {\bibfnamefont {F.~J.}\ \bibnamefont {Burnell}},\
  }\bibfield  {title} {\enquote {\bibinfo {title} {Fractonic chern-simons and
  bf theories},}\ }\href {\doibase 10.1103/physrevresearch.2.023249} {\bibfield
   {journal} {\bibinfo  {journal} {Physical Review Research}\ }\textbf
  {\bibinfo {volume} {2}} (\bibinfo {year} {2020}{\natexlab{b}}),\
  10.1103/physrevresearch.2.023249}\BibitemShut {NoStop}%
\bibitem [{\citenamefont {Weinstein}\ \emph {et~al.}(2020)\citenamefont
  {Weinstein}, \citenamefont {Cobanera}, \citenamefont {Ortiz},\ and\
  \citenamefont {Nussinov}}]{absence_Weinstein}%
  \BibitemOpen
  \bibfield  {author} {\bibinfo {author} {\bibfnamefont {Zack}\ \bibnamefont
  {Weinstein}}, \bibinfo {author} {\bibfnamefont {Emilio}\ \bibnamefont
  {Cobanera}}, \bibinfo {author} {\bibfnamefont {Gerardo}\ \bibnamefont
  {Ortiz}}, \ and\ \bibinfo {author} {\bibfnamefont {Zohar}\ \bibnamefont
  {Nussinov}},\ }\bibfield  {title} {\enquote {\bibinfo {title} {Absence of
  finite temperature phase transitions in the x-cube model and its zp
  generalization},}\ }\href {\doibase 10.1016/j.aop.2019.168018} {\bibfield
  {journal} {\bibinfo  {journal} {Annals of Physics}\ }\textbf {\bibinfo
  {volume} {412}},\ \bibinfo {pages} {168018} (\bibinfo {year}
  {2020})}\BibitemShut {NoStop}%
\bibitem [{\citenamefont {Wang}\ \emph {et~al.}(2021)\citenamefont {Wang},
  \citenamefont {Xu},\ and\ \citenamefont {Yau}}]{higher_Wang}%
  \BibitemOpen
  \bibfield  {author} {\bibinfo {author} {\bibfnamefont {Juven}\ \bibnamefont
  {Wang}}, \bibinfo {author} {\bibfnamefont {Kai}\ \bibnamefont {Xu}}, \ and\
  \bibinfo {author} {\bibfnamefont {Shing-Tung}\ \bibnamefont {Yau}},\
  }\bibfield  {title} {\enquote {\bibinfo {title} {Higher-rank tensor
  non-abelian field theory: Higher-moment or subdimensional polynomial global
  symmetry, algebraic variety, noether’s theorem, and gauging},}\ }\href
  {\doibase 10.1103/physrevresearch.3.013185} {\bibfield  {journal} {\bibinfo
  {journal} {Physical Review Research}\ }\textbf {\bibinfo {volume} {3}}
  (\bibinfo {year} {2021}),\ 10.1103/physrevresearch.3.013185}\BibitemShut
  {NoStop}%
\bibitem [{\citenamefont {Seiberg}(2020)}]{field_Seiberg}%
  \BibitemOpen
  \bibfield  {author} {\bibinfo {author} {\bibfnamefont {Nathan}\ \bibnamefont
  {Seiberg}},\ }\bibfield  {title} {\enquote {\bibinfo {title} {Field theories
  with a vector global symmetry},}\ }\href {\doibase
  10.21468/scipostphys.8.4.050} {\bibfield  {journal} {\bibinfo  {journal}
  {SciPost Physics}\ }\textbf {\bibinfo {volume} {8}} (\bibinfo {year}
  {2020}),\ 10.21468/scipostphys.8.4.050}\BibitemShut {NoStop}%
\bibitem [{\citenamefont {Aasen}\ \emph {et~al.}(2020)\citenamefont {Aasen},
  \citenamefont {Bulmash}, \citenamefont {Prem}, \citenamefont {Slagle},\ and\
  \citenamefont {Williamson}}]{topological_Aasen}%
  \BibitemOpen
  \bibfield  {author} {\bibinfo {author} {\bibfnamefont {David}\ \bibnamefont
  {Aasen}}, \bibinfo {author} {\bibfnamefont {Daniel}\ \bibnamefont {Bulmash}},
  \bibinfo {author} {\bibfnamefont {Abhinav}\ \bibnamefont {Prem}}, \bibinfo
  {author} {\bibfnamefont {Kevin}\ \bibnamefont {Slagle}}, \ and\ \bibinfo
  {author} {\bibfnamefont {Dominic~J.}\ \bibnamefont {Williamson}},\ }\bibfield
   {title} {\enquote {\bibinfo {title} {Topological defect networks for
  fractons of all types},}\ }\href {\doibase 10.1103/physrevresearch.2.043165}
  {\bibfield  {journal} {\bibinfo  {journal} {Physical Review Research}\
  }\textbf {\bibinfo {volume} {2}} (\bibinfo {year} {2020}),\
  10.1103/physrevresearch.2.043165}\BibitemShut {NoStop}%
\bibitem [{\citenamefont {Ma}\ \emph {et~al.}(2017)\citenamefont {Ma},
  \citenamefont {Lake}, \citenamefont {Chen},\ and\ \citenamefont
  {Hermele}}]{fracton_Hermele}%
  \BibitemOpen
  \bibfield  {author} {\bibinfo {author} {\bibfnamefont {Han}\ \bibnamefont
  {Ma}}, \bibinfo {author} {\bibfnamefont {Ethan}\ \bibnamefont {Lake}},
  \bibinfo {author} {\bibfnamefont {Xie}\ \bibnamefont {Chen}}, \ and\ \bibinfo
  {author} {\bibfnamefont {Michael}\ \bibnamefont {Hermele}},\ }\bibfield
  {title} {\enquote {\bibinfo {title} {Fracton topological order via coupled
  layers},}\ }\href {\doibase 10.1103/PhysRevB.95.245126} {\bibfield  {journal}
  {\bibinfo  {journal} {Phys. Rev. B}\ }\textbf {\bibinfo {volume} {95}},\
  \bibinfo {pages} {245126} (\bibinfo {year} {2017})}\BibitemShut {NoStop}%
\bibitem [{\citenamefont {Yuan}\ \emph {et~al.}(2020)\citenamefont {Yuan},
  \citenamefont {Chen},\ and\ \citenamefont {Ye}}]{fractonic_Ye}%
  \BibitemOpen
  \bibfield  {author} {\bibinfo {author} {\bibfnamefont {Jian-Keng}\
  \bibnamefont {Yuan}}, \bibinfo {author} {\bibfnamefont {Shuai~A.}\
  \bibnamefont {Chen}}, \ and\ \bibinfo {author} {\bibfnamefont {Peng}\
  \bibnamefont {Ye}},\ }\bibfield  {title} {\enquote {\bibinfo {title}
  {Fractonic superfluids},}\ }\href {\doibase 10.1103/physrevresearch.2.023267}
  {\bibfield  {journal} {\bibinfo  {journal} {Physical Review Research}\
  }\textbf {\bibinfo {volume} {2}} (\bibinfo {year} {2020}),\
  10.1103/physrevresearch.2.023267}\BibitemShut {NoStop}%
\bibitem [{\citenamefont {Ma}\ \emph {et~al.}(2018)\citenamefont {Ma},
  \citenamefont {Schmitz}, \citenamefont {Parameswaran}, \citenamefont
  {Hermele},\ and\ \citenamefont {Nandkishore}}]{topological_Nandkishore}%
  \BibitemOpen
  \bibfield  {author} {\bibinfo {author} {\bibfnamefont {Han}\ \bibnamefont
  {Ma}}, \bibinfo {author} {\bibfnamefont {A.~T.}\ \bibnamefont {Schmitz}},
  \bibinfo {author} {\bibfnamefont {S.~A.}\ \bibnamefont {Parameswaran}},
  \bibinfo {author} {\bibfnamefont {Michael}\ \bibnamefont {Hermele}}, \ and\
  \bibinfo {author} {\bibfnamefont {Rahul~M.}\ \bibnamefont {Nandkishore}},\
  }\bibfield  {title} {\enquote {\bibinfo {title} {Topological entanglement
  entropy of fracton stabilizer codes},}\ }\href {\doibase
  10.1103/physrevb.97.125101} {\bibfield  {journal} {\bibinfo  {journal}
  {Physical Review B}\ }\textbf {\bibinfo {volume} {97}} (\bibinfo {year}
  {2018}),\ 10.1103/physrevb.97.125101}\BibitemShut {NoStop}%
\bibitem [{\citenamefont {Schmitz}\ \emph {et~al.}(2017)\citenamefont
  {Schmitz}, \citenamefont {Ma},\ and\ \citenamefont
  {Nandkishore}}]{recoverable_Nandkishore}%
  \BibitemOpen
  \bibfield  {author} {\bibinfo {author} {\bibfnamefont {A.}~\bibnamefont
  {Schmitz}}, \bibinfo {author} {\bibfnamefont {Han}\ \bibnamefont {Ma}}, \
  and\ \bibinfo {author} {\bibfnamefont {Rahul}\ \bibnamefont {Nandkishore}},\
  }\bibfield  {title} {\enquote {\bibinfo {title} {Recoverable information and
  emergent conservation laws in fracton stabilizer codes},}\ }\href {\doibase
  10.1103/PhysRevB.97.134426} {\bibfield  {journal} {\bibinfo  {journal}
  {Physical Review B}\ }\textbf {\bibinfo {volume} {97}} (\bibinfo {year}
  {2017}),\ 10.1103/PhysRevB.97.134426}\BibitemShut {NoStop}%
\bibitem [{\citenamefont {Ma}\ and\ \citenamefont
  {Pretko}(2018)}]{higher_Pretko_2}%
  \BibitemOpen
  \bibfield  {author} {\bibinfo {author} {\bibfnamefont {Han}\ \bibnamefont
  {Ma}}\ and\ \bibinfo {author} {\bibfnamefont {Michael}\ \bibnamefont
  {Pretko}},\ }\bibfield  {title} {\enquote {\bibinfo {title} {Higher-rank
  deconfined quantum criticality at the lifshitz transition and the exciton
  bose condensate},}\ }\href {\doibase 10.1103/PhysRevB.98.125105} {\bibfield
  {journal} {\bibinfo  {journal} {Phys. Rev. B}\ }\textbf {\bibinfo {volume}
  {98}},\ \bibinfo {pages} {125105} (\bibinfo {year} {2018})}\BibitemShut
  {NoStop}%
\bibitem [{\citenamefont {Moudgalya}\ \emph {et~al.}(2019)\citenamefont
  {Moudgalya}, \citenamefont {Prem}, \citenamefont {Nandkishore}, \citenamefont
  {Regnault},\ and\ \citenamefont {Bernevig}}]{thermalization_Bernevig}%
  \BibitemOpen
  \bibfield  {author} {\bibinfo {author} {\bibfnamefont {Sanjay}\ \bibnamefont
  {Moudgalya}}, \bibinfo {author} {\bibfnamefont {Abhinav}\ \bibnamefont
  {Prem}}, \bibinfo {author} {\bibfnamefont {Rahul}\ \bibnamefont
  {Nandkishore}}, \bibinfo {author} {\bibfnamefont {Nicolas}\ \bibnamefont
  {Regnault}}, \ and\ \bibinfo {author} {\bibfnamefont {B.~Andrei}\
  \bibnamefont {Bernevig}},\ }\href@noop {} {\enquote {\bibinfo {title}
  {Thermalization and its absence within krylov subspaces of a constrained
  hamiltonian},}\ } (\bibinfo {year} {2019}),\ \Eprint
  {http://arxiv.org/abs/1910.14048} {arXiv:1910.14048 [cond-mat.str-el]}
  \BibitemShut {NoStop}%
\bibitem [{\citenamefont {Sous}\ and\ \citenamefont
  {Pretko}(2020)}]{fractons_sous}%
  \BibitemOpen
  \bibfield  {author} {\bibinfo {author} {\bibfnamefont {John}\ \bibnamefont
  {Sous}}\ and\ \bibinfo {author} {\bibfnamefont {Michael}\ \bibnamefont
  {Pretko}},\ }\bibfield  {title} {\enquote {\bibinfo {title} {Fractons from
  polarons},}\ }\href {\doibase 10.1103/physrevb.102.214437} {\bibfield
  {journal} {\bibinfo  {journal} {Physical Review B}\ }\textbf {\bibinfo
  {volume} {102}} (\bibinfo {year} {2020}),\
  10.1103/physrevb.102.214437}\BibitemShut {NoStop}%
\bibitem [{\citenamefont {Gromov}\ \emph {et~al.}(2020)\citenamefont {Gromov},
  \citenamefont {Lucas},\ and\ \citenamefont {Nandkishore}}]{fracton_hydro}%
  \BibitemOpen
  \bibfield  {author} {\bibinfo {author} {\bibfnamefont {Andrey}\ \bibnamefont
  {Gromov}}, \bibinfo {author} {\bibfnamefont {Andrew}\ \bibnamefont {Lucas}},
  \ and\ \bibinfo {author} {\bibfnamefont {Rahul~M.}\ \bibnamefont
  {Nandkishore}},\ }\bibfield  {title} {\enquote {\bibinfo {title} {Fracton
  hydrodynamics},}\ }\href {\doibase 10.1103/PhysRevResearch.2.033124}
  {\bibfield  {journal} {\bibinfo  {journal} {Phys. Rev. Research}\ }\textbf
  {\bibinfo {volume} {2}},\ \bibinfo {pages} {033124} (\bibinfo {year}
  {2020})}\BibitemShut {NoStop}%
\bibitem [{\citenamefont {Feldmeier}\ \emph {et~al.}(2020)\citenamefont
  {Feldmeier}, \citenamefont {Sala}, \citenamefont {De~Tomasi}, \citenamefont
  {Pollmann},\ and\ \citenamefont {Knap}}]{knap2020}%
  \BibitemOpen
  \bibfield  {author} {\bibinfo {author} {\bibfnamefont {Johannes}\
  \bibnamefont {Feldmeier}}, \bibinfo {author} {\bibfnamefont {Pablo}\
  \bibnamefont {Sala}}, \bibinfo {author} {\bibfnamefont {Giuseppe}\
  \bibnamefont {De~Tomasi}}, \bibinfo {author} {\bibfnamefont {Frank}\
  \bibnamefont {Pollmann}}, \ and\ \bibinfo {author} {\bibfnamefont {Michael}\
  \bibnamefont {Knap}},\ }\bibfield  {title} {\enquote {\bibinfo {title}
  {Anomalous diffusion in dipole- and higher-moment-conserving systems},}\
  }\href {\doibase 10.1103/PhysRevLett.125.245303} {\bibfield  {journal}
  {\bibinfo  {journal} {Phys. Rev. Lett.}\ }\textbf {\bibinfo {volume} {125}},\
  \bibinfo {pages} {245303} (\bibinfo {year} {2020})}\BibitemShut {NoStop}%
\bibitem [{\citenamefont {Morningstar}\ \emph {et~al.}(2020)\citenamefont
  {Morningstar}, \citenamefont {Khemani},\ and\ \citenamefont
  {Huse}}]{morningstar}%
  \BibitemOpen
  \bibfield  {author} {\bibinfo {author} {\bibfnamefont {Alan}\ \bibnamefont
  {Morningstar}}, \bibinfo {author} {\bibfnamefont {Vedika}\ \bibnamefont
  {Khemani}}, \ and\ \bibinfo {author} {\bibfnamefont {David~A.}\ \bibnamefont
  {Huse}},\ }\bibfield  {title} {\enquote {\bibinfo {title} {Kinetically
  constrained freezing transition in a dipole-conserving system},}\ }\href
  {\doibase 10.1103/PhysRevB.101.214205} {\bibfield  {journal} {\bibinfo
  {journal} {Phys. Rev. B}\ }\textbf {\bibinfo {volume} {101}},\ \bibinfo
  {pages} {214205} (\bibinfo {year} {2020})}\BibitemShut {NoStop}%
\bibitem [{\citenamefont {Iaconis}\ \emph {et~al.}(2021)\citenamefont
  {Iaconis}, \citenamefont {Lucas},\ and\ \citenamefont
  {Nandkishore}}]{conservation_Nandkishore}%
  \BibitemOpen
  \bibfield  {author} {\bibinfo {author} {\bibfnamefont {J.}~\bibnamefont
  {Iaconis}}, \bibinfo {author} {\bibfnamefont {A.}~\bibnamefont {Lucas}}, \
  and\ \bibinfo {author} {\bibfnamefont {R.}~\bibnamefont {Nandkishore}},\
  }\bibfield  {title} {\enquote {\bibinfo {title} {Multipole conservation laws
  and subdiffusion in any dimension.}}\ }\href@noop {} {\bibfield  {journal}
  {\bibinfo  {journal} {Physical review. E}\ }\textbf {\bibinfo {volume} {103
  2-1}},\ \bibinfo {pages} {022142} (\bibinfo {year} {2021})}\BibitemShut
  {NoStop}%
\bibitem [{\citenamefont {Hart}\ \emph {et~al.}(2021)\citenamefont {Hart},
  \citenamefont {Lucas},\ and\ \citenamefont {Nandkishore}}]{hart2021hidden}%
  \BibitemOpen
  \bibfield  {author} {\bibinfo {author} {\bibfnamefont {Oliver}\ \bibnamefont
  {Hart}}, \bibinfo {author} {\bibfnamefont {Andrew}\ \bibnamefont {Lucas}}, \
  and\ \bibinfo {author} {\bibfnamefont {Rahul}\ \bibnamefont {Nandkishore}},\
  }\href@noop {} {\enquote {\bibinfo {title} {Hidden quasi-conservation laws in
  fracton hydrodynamics},}\ } (\bibinfo {year} {2021}),\ \Eprint
  {http://arxiv.org/abs/2110.08292} {arXiv:2110.08292 [cond-mat.stat-mech]}
  \BibitemShut {NoStop}%
\bibitem [{\citenamefont {Sala}\ \emph {et~al.}(2021)\citenamefont {Sala},
  \citenamefont {Lehmann}, \citenamefont {Rakovszky},\ and\ \citenamefont
  {Pollmann}}]{sala2021dynamics}%
  \BibitemOpen
  \bibfield  {author} {\bibinfo {author} {\bibfnamefont {Pablo}\ \bibnamefont
  {Sala}}, \bibinfo {author} {\bibfnamefont {Julius}\ \bibnamefont {Lehmann}},
  \bibinfo {author} {\bibfnamefont {Tibor}\ \bibnamefont {Rakovszky}}, \ and\
  \bibinfo {author} {\bibfnamefont {Frank}\ \bibnamefont {Pollmann}},\
  }\href@noop {} {\enquote {\bibinfo {title} {Dynamics in systems with
  modulated symmetries},}\ } (\bibinfo {year} {2021}),\ \Eprint
  {http://arxiv.org/abs/2110.08302} {arXiv:2110.08302 [cond-mat.stat-mech]}
  \BibitemShut {NoStop}%
\bibitem [{\citenamefont {Iaconis}\ \emph {et~al.}(2019)\citenamefont
  {Iaconis}, \citenamefont {Vijay},\ and\ \citenamefont
  {Nandkishore}}]{IaconisVijayNandkishore}%
  \BibitemOpen
  \bibfield  {author} {\bibinfo {author} {\bibfnamefont {Jason}\ \bibnamefont
  {Iaconis}}, \bibinfo {author} {\bibfnamefont {Sagar}\ \bibnamefont {Vijay}},
  \ and\ \bibinfo {author} {\bibfnamefont {Rahul}\ \bibnamefont
  {Nandkishore}},\ }\bibfield  {title} {\enquote {\bibinfo {title} {Anomalous
  subdiffusion from subsystem symmetries},}\ }\href {\doibase
  10.1103/PhysRevB.100.214301} {\bibfield  {journal} {\bibinfo  {journal}
  {Phys. Rev. B}\ }\textbf {\bibinfo {volume} {100}},\ \bibinfo {pages}
  {214301} (\bibinfo {year} {2019})}\BibitemShut {NoStop}%
\bibitem [{\citenamefont {Feldmeier}\ \emph {et~al.}(2021)\citenamefont
  {Feldmeier}, \citenamefont {Pollmann},\ and\ \citenamefont
  {Knap}}]{knap2021}%
  \BibitemOpen
  \bibfield  {author} {\bibinfo {author} {\bibfnamefont {Johannes}\
  \bibnamefont {Feldmeier}}, \bibinfo {author} {\bibfnamefont {Frank}\
  \bibnamefont {Pollmann}}, \ and\ \bibinfo {author} {\bibfnamefont {Michael}\
  \bibnamefont {Knap}},\ }\bibfield  {title} {\enquote {\bibinfo {title}
  {Emergent fracton dynamics in a nonplanar dimer model},}\ }\href {\doibase
  10.1103/PhysRevB.103.094303} {\bibfield  {journal} {\bibinfo  {journal}
  {Phys. Rev. B}\ }\textbf {\bibinfo {volume} {103}},\ \bibinfo {pages}
  {094303} (\bibinfo {year} {2021})}\BibitemShut {NoStop}%
\bibitem [{\citenamefont {Guardado-Sanchez}\ \emph {et~al.}(2020)\citenamefont
  {Guardado-Sanchez}, \citenamefont {Morningstar}, \citenamefont {Spar},
  \citenamefont {Brown}, \citenamefont {Huse},\ and\ \citenamefont
  {Bakr}}]{Guardado_Sanchez_2020}%
  \BibitemOpen
  \bibfield  {author} {\bibinfo {author} {\bibfnamefont {Elmer}\ \bibnamefont
  {Guardado-Sanchez}}, \bibinfo {author} {\bibfnamefont {Alan}\ \bibnamefont
  {Morningstar}}, \bibinfo {author} {\bibfnamefont {Benjamin~M.}\ \bibnamefont
  {Spar}}, \bibinfo {author} {\bibfnamefont {Peter~T.}\ \bibnamefont {Brown}},
  \bibinfo {author} {\bibfnamefont {David~A.}\ \bibnamefont {Huse}}, \ and\
  \bibinfo {author} {\bibfnamefont {Waseem~S.}\ \bibnamefont {Bakr}},\
  }\bibfield  {title} {\enquote {\bibinfo {title} {{Subdiffusion and Heat
  Transport in a Tilted Two-Dimensional Fermi-Hubbard System}},}\ }\href
  {\doibase 10.1103/physrevx.10.011042} {\bibfield  {journal} {\bibinfo
  {journal} {Physical Review X}\ }\textbf {\bibinfo {volume} {10}},\ \bibinfo
  {pages} {011042} (\bibinfo {year} {2020})}\BibitemShut {NoStop}%
\bibitem [{\citenamefont {Zhang}(2020)}]{zhang2020universal}%
  \BibitemOpen
  \bibfield  {author} {\bibinfo {author} {\bibfnamefont {Pengfei}\ \bibnamefont
  {Zhang}},\ }\href@noop {} {\enquote {\bibinfo {title} {Universal subdiffusion
  in strongly tilted many-body systems},}\ } (\bibinfo {year} {2020}),\ \Eprint
  {http://arxiv.org/abs/2004.08695} {arXiv:2004.08695 [cond-mat.quant-gas]}
  \BibitemShut {NoStop}%
\bibitem [{\citenamefont {Glorioso}\ \emph {et~al.}(2021)\citenamefont
  {Glorioso}, \citenamefont {Guo}, \citenamefont {Rodriguez-Nieva},\ and\
  \citenamefont {Lucas}}]{lt4dbreakdown_Lucas}%
  \BibitemOpen
  \bibfield  {author} {\bibinfo {author} {\bibfnamefont {Paolo}\ \bibnamefont
  {Glorioso}}, \bibinfo {author} {\bibfnamefont {Jinkang}\ \bibnamefont {Guo}},
  \bibinfo {author} {\bibfnamefont {Joaquin~F.}\ \bibnamefont
  {Rodriguez-Nieva}}, \ and\ \bibinfo {author} {\bibfnamefont {Andrew}\
  \bibnamefont {Lucas}},\ }\href@noop {} {\enquote {\bibinfo {title} {Breakdown
  of hydrodynamics below four dimensions in a fracton fluid},}\ } (\bibinfo
  {year} {2021}),\ \Eprint {http://arxiv.org/abs/2105.13365} {arXiv:2105.13365
  [cond-mat.str-el]} \BibitemShut {NoStop}%
\bibitem [{\citenamefont {Grosvenor}\ \emph {et~al.}(2021)\citenamefont
  {Grosvenor}, \citenamefont {Hoyos}, \citenamefont {Peña-Benitez},\ and\
  \citenamefont {Surówka}}]{hydroscoop_Surowka}%
  \BibitemOpen
  \bibfield  {author} {\bibinfo {author} {\bibfnamefont {Kevin~T.}\
  \bibnamefont {Grosvenor}}, \bibinfo {author} {\bibfnamefont {Carlos}\
  \bibnamefont {Hoyos}}, \bibinfo {author} {\bibfnamefont {Francisco}\
  \bibnamefont {Peña-Benitez}}, \ and\ \bibinfo {author} {\bibfnamefont
  {Piotr}\ \bibnamefont {Surówka}},\ }\href@noop {} {\enquote {\bibinfo
  {title} {Hydrodynamics of ideal fracton fluids},}\ } (\bibinfo {year}
  {2021}),\ \Eprint {http://arxiv.org/abs/2105.01084} {arXiv:2105.01084
  [cond-mat.str-el]} \BibitemShut {NoStop}%
\bibitem [{\citenamefont {Slagle}\ \emph
  {et~al.}(2019{\natexlab{c}})\citenamefont {Slagle}, \citenamefont {Prem},\
  and\ \citenamefont {Pretko}}]{Slagle:2018kqf}%
  \BibitemOpen
  \bibfield  {author} {\bibinfo {author} {\bibfnamefont {Kevin}\ \bibnamefont
  {Slagle}}, \bibinfo {author} {\bibfnamefont {Abhinav}\ \bibnamefont {Prem}},
  \ and\ \bibinfo {author} {\bibfnamefont {Michael}\ \bibnamefont {Pretko}},\
  }\bibfield  {title} {\enquote {\bibinfo {title} {{Symmetric Tensor Gauge
  Theories on Curved Spaces}},}\ }\href {\doibase 10.1016/j.aop.2019.167910}
  {\bibfield  {journal} {\bibinfo  {journal} {Annals Phys.}\ }\textbf {\bibinfo
  {volume} {410}},\ \bibinfo {pages} {167910} (\bibinfo {year}
  {2019}{\natexlab{c}})},\ \Eprint {http://arxiv.org/abs/1807.00827}
  {arXiv:1807.00827 [cond-mat.str-el]} \BibitemShut {NoStop}%
\bibitem [{\citenamefont {Crossley}\ \emph {et~al.}(2015)\citenamefont
  {Crossley}, \citenamefont {Glorioso},\ and\ \citenamefont {Liu}}]{EFT_Liu}%
  \BibitemOpen
  \bibfield  {author} {\bibinfo {author} {\bibfnamefont {Michael}\ \bibnamefont
  {Crossley}}, \bibinfo {author} {\bibfnamefont {Paolo}\ \bibnamefont
  {Glorioso}}, \ and\ \bibinfo {author} {\bibfnamefont {Hong}\ \bibnamefont
  {Liu}},\ }\bibfield  {title} {\enquote {\bibinfo {title} {Effective field
  theory of dissipative fluids},}\ }\href {\doibase 10.1007/JHEP09(2017)095}
  {\bibfield  {journal} {\bibinfo  {journal} {Journal of High Energy Physics}\
  }\textbf {\bibinfo {volume} {2017}} (\bibinfo {year} {2015}),\
  10.1007/JHEP09(2017)095}\BibitemShut {NoStop}%
\bibitem [{\citenamefont {Glorioso}\ \emph {et~al.}()\citenamefont {Glorioso},
  \citenamefont {Guo}, \citenamefont {Rodriguez-Nieva},\ and\ \citenamefont
  {Lucas}}]{paolotoappear}%
  \BibitemOpen
  \bibfield  {author} {\bibinfo {author} {\bibfnamefont {Paolo}\ \bibnamefont
  {Glorioso}}, \bibinfo {author} {\bibfnamefont {Jinkang}\ \bibnamefont {Guo}},
  \bibinfo {author} {\bibfnamefont {Joaquin~F.}\ \bibnamefont
  {Rodriguez-Nieva}}, \ and\ \bibinfo {author} {\bibfnamefont {Andrew}\
  \bibnamefont {Lucas}},\ }\href@noop {} {\enquote {\bibinfo {title} {to
  appear},}\ }\BibitemShut {NoStop}%
\bibitem [{\citenamefont {Jain}\ and\ \citenamefont
  {Jensen}(2021)}]{Jain:2021ibh}%
  \BibitemOpen
  \bibfield  {author} {\bibinfo {author} {\bibfnamefont {Akash}\ \bibnamefont
  {Jain}}\ and\ \bibinfo {author} {\bibfnamefont {Kristan}\ \bibnamefont
  {Jensen}},\ }\bibfield  {title} {\enquote {\bibinfo {title} {{Fractons in
  curved space}},}\ }\href@noop {} {\  (\bibinfo {year} {2021})},\ \Eprint
  {http://arxiv.org/abs/2111.03973} {arXiv:2111.03973 [hep-th]} \BibitemShut
  {NoStop}%
\bibitem [{\citenamefont {Bidussi}\ \emph {et~al.}(2021)\citenamefont
  {Bidussi}, \citenamefont {Hartong}, \citenamefont {Have}, \citenamefont
  {Musaeus},\ and\ \citenamefont {Prohazka}}]{Bidussi:2021nmp}%
  \BibitemOpen
  \bibfield  {author} {\bibinfo {author} {\bibfnamefont {Leo}\ \bibnamefont
  {Bidussi}}, \bibinfo {author} {\bibfnamefont {Jelle}\ \bibnamefont
  {Hartong}}, \bibinfo {author} {\bibfnamefont {Emil}\ \bibnamefont {Have}},
  \bibinfo {author} {\bibfnamefont {J\o{}rgen}\ \bibnamefont {Musaeus}}, \ and\
  \bibinfo {author} {\bibfnamefont {Stefan}\ \bibnamefont {Prohazka}},\
  }\bibfield  {title} {\enquote {\bibinfo {title} {{Fractons, dipole symmetries
  and curved spacetime}},}\ }\href@noop {} {\  (\bibinfo {year} {2021})},\
  \Eprint {http://arxiv.org/abs/2111.03668} {arXiv:2111.03668 [hep-th]}
  \BibitemShut {NoStop}%
\bibitem [{\citenamefont {Pe\~na Benitez}(2021)}]{Pena-Benitez:2021ipo}%
  \BibitemOpen
  \bibfield  {author} {\bibinfo {author} {\bibfnamefont {Francisco}\
  \bibnamefont {Pe\~na Benitez}},\ }\bibfield  {title} {\enquote {\bibinfo
  {title} {{Fractons, Symmetric Gauge Fields and Geometry}},}\ }\href@noop {}
  {\  (\bibinfo {year} {2021})},\ \Eprint {http://arxiv.org/abs/2107.13884}
  {arXiv:2107.13884 [cond-mat.str-el]} \BibitemShut {NoStop}%
\bibitem [{\citenamefont {Son}(2001)}]{Son:2000ht}%
  \BibitemOpen
  \bibfield  {author} {\bibinfo {author} {\bibfnamefont {D.~T.}\ \bibnamefont
  {Son}},\ }\bibfield  {title} {\enquote {\bibinfo {title} {{Hydrodynamics of
  relativistic systems with broken continuous symmetries}},}\ }\href {\doibase
  10.1142/S0217751X01009545} {\bibfield  {journal} {\bibinfo  {journal} {Int.
  J. Mod. Phys. A}\ }\textbf {\bibinfo {volume} {16S1C}},\ \bibinfo {pages}
  {1284--1286} (\bibinfo {year} {2001})},\ \Eprint
  {http://arxiv.org/abs/hep-ph/0011246} {arXiv:hep-ph/0011246} \BibitemShut
  {NoStop}%
\bibitem [{\citenamefont {Gromov}(2019)}]{multipole_gromov}%
  \BibitemOpen
  \bibfield  {author} {\bibinfo {author} {\bibfnamefont {Andrey}\ \bibnamefont
  {Gromov}},\ }\bibfield  {title} {\enquote {\bibinfo {title} {Towards
  classification of fracton phases: The multipole algebra},}\ }\href {\doibase
  10.1103/PhysRevX.9.031035} {\bibfield  {journal} {\bibinfo  {journal}
  {Physical Review X}\ }\textbf {\bibinfo {volume} {9}} (\bibinfo {year}
  {2019}),\ 10.1103/PhysRevX.9.031035}\BibitemShut {NoStop}%
\bibitem [{\citenamefont {Kardar}\ \emph {et~al.}(1986)\citenamefont {Kardar},
  \citenamefont {Parisi},\ and\ \citenamefont {Zhang}}]{kpz}%
  \BibitemOpen
  \bibfield  {author} {\bibinfo {author} {\bibfnamefont {Mehran}\ \bibnamefont
  {Kardar}}, \bibinfo {author} {\bibfnamefont {Giorgio}\ \bibnamefont
  {Parisi}}, \ and\ \bibinfo {author} {\bibfnamefont {Yi-Cheng}\ \bibnamefont
  {Zhang}},\ }\bibfield  {title} {\enquote {\bibinfo {title} {Dynamic scaling
  of growing interfaces},}\ }\href {\doibase 10.1103/PhysRevLett.56.889}
  {\bibfield  {journal} {\bibinfo  {journal} {Phys. Rev. Lett.}\ }\textbf
  {\bibinfo {volume} {56}},\ \bibinfo {pages} {889--892} (\bibinfo {year}
  {1986})}\BibitemShut {NoStop}%
\bibitem [{\citenamefont {Das}\ \emph {et~al.}(2014)\citenamefont {Das},
  \citenamefont {Dhar}, \citenamefont {Saito}, \citenamefont {Mendl},\ and\
  \citenamefont {Spohn}}]{PhysRevE.90.012124}%
  \BibitemOpen
  \bibfield  {author} {\bibinfo {author} {\bibfnamefont {Suman~G.}\
  \bibnamefont {Das}}, \bibinfo {author} {\bibfnamefont {Abhishek}\
  \bibnamefont {Dhar}}, \bibinfo {author} {\bibfnamefont {Keiji}\ \bibnamefont
  {Saito}}, \bibinfo {author} {\bibfnamefont {Christian~B.}\ \bibnamefont
  {Mendl}}, \ and\ \bibinfo {author} {\bibfnamefont {Herbert}\ \bibnamefont
  {Spohn}},\ }\bibfield  {title} {\enquote {\bibinfo {title} {Numerical test of
  hydrodynamic fluctuation theory in the fermi-pasta-ulam chain},}\ }\href
  {\doibase 10.1103/PhysRevE.90.012124} {\bibfield  {journal} {\bibinfo
  {journal} {Phys. Rev. E}\ }\textbf {\bibinfo {volume} {90}},\ \bibinfo
  {pages} {012124} (\bibinfo {year} {2014})}\BibitemShut {NoStop}%
\bibitem [{\citenamefont {Grozdanov}\ \emph {et~al.}(2019)\citenamefont
  {Grozdanov}, \citenamefont {Lucas},\ and\ \citenamefont
  {Poovuttikul}}]{Grozdanov:2018fic}%
  \BibitemOpen
  \bibfield  {author} {\bibinfo {author} {\bibfnamefont {Sa\v{s}o}\
  \bibnamefont {Grozdanov}}, \bibinfo {author} {\bibfnamefont {Andrew}\
  \bibnamefont {Lucas}}, \ and\ \bibinfo {author} {\bibfnamefont {Napat}\
  \bibnamefont {Poovuttikul}},\ }\bibfield  {title} {\enquote {\bibinfo {title}
  {{Holography and hydrodynamics with weakly broken symmetries}},}\ }\href
  {\doibase 10.1103/PhysRevD.99.086012} {\bibfield  {journal} {\bibinfo
  {journal} {Phys. Rev. D}\ }\textbf {\bibinfo {volume} {99}},\ \bibinfo
  {pages} {086012} (\bibinfo {year} {2019})},\ \Eprint
  {http://arxiv.org/abs/1810.10016} {arXiv:1810.10016 [hep-th]} \BibitemShut
  {NoStop}%
\bibitem [{\citenamefont {Baggioli}\ \emph {et~al.}(2020)\citenamefont
  {Baggioli}, \citenamefont {Vasin}, \citenamefont {Brazhkin},\ and\
  \citenamefont {Trachenko}}]{Baggioli:2019jcm}%
  \BibitemOpen
  \bibfield  {author} {\bibinfo {author} {\bibfnamefont {Matteo}\ \bibnamefont
  {Baggioli}}, \bibinfo {author} {\bibfnamefont {Mikhail}\ \bibnamefont
  {Vasin}}, \bibinfo {author} {\bibfnamefont {Vadim~V.}\ \bibnamefont
  {Brazhkin}}, \ and\ \bibinfo {author} {\bibfnamefont {Kostya}\ \bibnamefont
  {Trachenko}},\ }\bibfield  {title} {\enquote {\bibinfo {title} {{Gapped
  momentum states}},}\ }\href {\doibase 10.1016/j.physrep.2020.04.002}
  {\bibfield  {journal} {\bibinfo  {journal} {Phys. Rept.}\ }\textbf {\bibinfo
  {volume} {865}},\ \bibinfo {pages} {1--44} (\bibinfo {year} {2020})},\
  \Eprint {http://arxiv.org/abs/1904.01419} {arXiv:1904.01419
  [cond-mat.stat-mech]} \BibitemShut {NoStop}%
\end{thebibliography}%
\end{document}